\documentclass[review]{elsarticle}

\usepackage{amssymb}

\let\counterwithin\relax
\usepackage{chngcntr}
\usepackage{color}
\usepackage{graphicx}
\usepackage{dcolumn}
\usepackage{bm}
\usepackage{mathrsfs}
\usepackage[ruled]{algorithm2e}
\usepackage{array} %
\usepackage[utf8]{inputenc}
\usepackage{tabstackengine}
\setstackEOL{\cr}
\usepackage{amsfonts,amsmath,mathrsfs,booktabs}
\usepackage{multirow,times}
\usepackage{float,color,bm,subfigure}
\usepackage{empheq}
\usepackage{hhline}
\usepackage{tikz}
\usepackage{lipsum,mathtools}
\usepackage[colorlinks,
linkcolor=blue,
anchorcolor=blue,
urlcolor=blue,
citecolor=blue
]{hyperref}

\newcommand{\splitatcommas}[1]{%
	\begingroup
	\ifnum\mathcode`,="8000
	\else
	\begingroup\lccode`~=`, \lowercase{\endgroup
		\edef~{\mathchar\the\mathcode`, \penalty0 \noexpand\hspace{0pt plus 0.3em}}%
	}\mathcode`,="8000
	\fi
	#1%
	\endgroup
}
\newcommand{\tuple}[1]{(\splitatcommas{#1})}
\newcommand{\set}[1]{\{\splitatcommas{#1}\}}
\newcommand{\RNum}[1]{\uppercase\expandafter{\romannumeral #1\relax}} 
\newcommand{\Rnum}[1]{\romannumeral #1\relax}
\newcommand{\PreserveBackslash}[1]{\let\temp=\\#1\let\\=\temp}
\newcolumntype{C}[1]{>{\PreserveBackslash\centering}p{#1}}
\newcolumntype{R}[1]{>{\PreserveBackslash\raggedleft}p{#1}}
\newcolumntype{L}[1]{>{\PreserveBackslash\raggedright}p{#1}}
 \normalsize

\journal{Chaos, Solitons $\&$ Fractals}

\begin{document}

\begin{frontmatter}



\title{Emergence of Cooperation in Two-agent Repeated Games with Reinforcement Learning}




\author[1]{Zhen-Wei Ding}
\author[2]{Guo-Zhong Zheng}
\author[3]{Chao-Ran Cai}
\author[4]{Wei-Ran Cai}
\author[2]{Li Chen}
\author[1]{Ji-Qiang Zhang\corref{cor1}}
\ead{zhangjq13@lzu.edu.cn}
\cortext[cor1]{Corresponding Author}
\author[1]{Xu-Ming Wang\corref{cor2}}
\ead{wang_xm@126.com}
\cortext[cor2]{Corresponding Author}
\address[1]{School of Physics, Ningxia University, Yinchuan 750021, P. R. China}
\address[2]{School of Physics and Information Technology, Shaanxi Normal University, Xi'an, 710062, P. R. China}
\address[3]{School of Physics, Northwest University, Xi’an 710127, P. R. China}
\address[4]{School of Computer Science, Soochow University, Suzhou 215006, P. R. China}

\begin{abstract}
Cooperation is the foundation of ecosystems and the human society, and the reinforcement learning provides crucial insight into the mechanism for its emergence. However, most previous work has mostly focused on the self-organization at the population level, the fundamental dynamics at the individual level remains unclear. Here, we investigate the evolution of cooperation in a two-agent system, where each agent pursues optimal policies according to the classical Q-learning algorithm in playing the strict prisoner's dilemma. 
We reveal that a strong memory and long-sighted expectation yield the emergence of \emph{Coordinated Optimal Policies} (COPs), where both agents act like ``Win-Stay, Lose-Shift'' (WSLS) to maintain a high level of cooperation. Otherwise, players become tolerant toward their co-player's defection and the cooperation loses stability in the end where the policy ``all Defection'' (All-D) prevails. This suggests that tolerance could be a good precursor to a crisis in cooperation. Furthermore, our analysis shows that the \emph{Coordinated Optimal Modes} (COMs) for different COPs gradually lose stability as memory weakens and expectation for the future decreases, where agents fail to predict co-player's action in games and defection dominates. 
As a result, we give the constraint to expectations of future and memory strength for maintaining cooperation. In contrast to the previous work, the impact of exploration on cooperation is found not be consistent, but depends on composition of COMs. By clarifying these fundamental issues in this two-player system, we hope that our work could be helpful for understanding the emergence and stability of cooperation in more complex scenarios in reality. 
\end{abstract}


\begin{highlights}
\item Strong memory and long-sighted expectations yield ``win-stay, lose-shift'' and high cooperation 
\item Tolerance of exploitation could be a good precursor to a crisis in cooperation
\item The impact of exploration on cooperation nonmonotonically depends on the composition of the coordinated optimal modes
\end{highlights}



\end{frontmatter}


\section{Introduction} 

Cooperation is ubiquitous and significant, from ant fortress associations and altruistic behavior of pathogenic bacteria in biological systems~\cite{bernasconi1999cooperation,griffin2004cooperation} to community activities and civic participation in human society~\cite{van2002introduction}. 
However, the emergence of cooperation is not straightforward, since although common interest would require the majority to cooperate, exploiting others by defection could maximize individuals' interest. Such dilemma arising from the conflict between collective and individual welfare is captured in a number of classical game models~\cite{rapoport1965prisoner, sachs2004evolution}. Here, the key question of how do cooperative behaviors evolve still remains an open question. 

Among the most favorable models for studying cooperation mechanisms, the Prisoner's Dilemma Game (PDG) stands out for its simplicity. It's well known that defection, as the Nash equilibrium, is the optimal choice for individuals in a single round for this game~\cite{nas1951non,S:1982}. But the repeated PDG potentially provides an escape to cooperation revealed in both theoretic predictions and experiments~\cite{luce1989games,murnighan1983expecting,kreps1982rational,axelrod1984evolution, kraines1993learning,nowak1993strategy,milinski1987tit,nowak1992tit}.
Previous studies show that the equilibriums of repeated PDGs depend crucially on whether a game is finitely or infinitely repeated~\cite{murnighan1983expecting,roth1978equilibrium}. 
There are some exceptions, however, that incomplete information, e.g. uncertain preferences of the players~\cite{kreps1982rational,andreoni1993rational}, uncertain number of rounds~\cite{van2011shadow,bo2005cooperation,camera2009cooperation}, termination rule~\cite{normann2012impact} or rewards with noise~\cite{bereby2006speed}, etc., can lead to the altruistic cooperation even in the finite repeated PDGs. 
This leads another interesting theme in repeated PDGs about the relevance of strategies to the cooperation level.  
Researchers have uncovered a number of strategies for actions, in which individuals' future deeds adhere to particular rules based on scant historical data, and they have investigated the dependence of the cooperation level on these rules~\cite{axelrod1981evolution,kreps1982rational,axelrod1984evolution, kraines1993learning,nowak1993strategy,milinski1987tit,nowak1992tit} as well as the choices made by the individuals within these strategies for actions~\cite{hilbe2015partners,dal2019strategy,wu2014boosting}. 
Based on the above works and the introduction of framework of evolutionary game theory, considerable progresses have been made later on around the mechanism behind the cooperation emergence among unrelated individuals~\cite{S:1982,wu2014boosting,perc2017statistical,deng2016self,hilbe2018partners,li2023investigating,zhu2023effects}.

It is notable that in most of these previous studies, the
strategies are not evolved or fixed once adopted, showing
weak adaptivity towards the circumstance.
Humans and many other creatures, however, have a sophisticated cognitive capacity, such as reinforcement  learning~\cite{bucsoniu2010multi}, behaviour prediction~\cite{devaine2014theory}, intention recognition~\cite{han2015synergy,han2011intention}, and intelligence from social interactions~\cite{mcnally2012cooperation}.  A new paradigm accounting for the adaptivity is needed to understand the cooperative behaviours in
reality.

The past decades have witnessed the flourishing of machine learning, which has rooted in human cognition and neuroscience~\cite{lee2008game,subramanian2022reinforcement} and has many successful applications in many fields~\cite{MCM:2013,YYG:2015,N:2007,TJLB:2014,CW:2006,silver2016mastering}. 
Reinforcement learning, as one of most influential branches of machine learning, is found particular suitable for understanding the evolution cooperation~\cite{masuda2011numerical,usui2021symmetric,horita2017reinforcement}. 
Reinforcement learning is originally designed to make optimal decision for maximizing the rewards for the given states through exploratory experimentation~\cite{subramanian2022reinforcement,kaelbling1996reinforcement,DTJIM:2018,AM:2003,WD:1992,HAD:2016,VKDAJ:2015}, and this setup exactly matches with the evolution of cooperation.
Actually, some studies have already adopted the idea of the reinforcement learning to investigate the repeated PDGs~\cite{sandholm1996multiagent,wunder2010classes,carmel1999exploration,harper2017reinforcement,kies2020finding,meylahn2022limiting,barfuss2023intrinsic}. 

With this new paradigm, new insights are obtained by studying the impact of different factors on cooperation~\cite{soton266919,barfuss2023intrinsic,xue2017adaptive}, e.g., it's found that cooperation can benefit from improved exploration methods~\cite{carmel1999exploration}, self-adaptive memories~\cite{xue2017adaptive}, evolved payoffs~\cite{vassiliades2010multiagent} and even intrinsic fluctuations~\cite{barfuss2023intrinsic}.
Some other works discuss the optimization of algorithms to facilitate the cooperation and increase rewards~\cite{kies2020finding,barnett2022oases,moriyama2009utility}, or find strategies to play against the classical strategies~\cite{sandholm1996multiagent,harper2017reinforcement}. 
In parallel, some theories have also been developed, such as symmetric equilibrium~\cite{usui2021symmetric}, symmetry breaking~\cite{fujimoto2019emergence} or fundamental dynamics~\cite{wunder2010classes,meylahn2022limiting,barfuss2019deterministic}. 
Building on these studies, researchers have
inspected cooperation from self-organization, in populations or multi-agent systems~\cite{oroojlooy2023review,yang2020overview,zhang2021multi}, which aim to continuously adjust strategies by learning instead of some fixed strategies, such as imitation learning in the classical evolutionary games~\cite{jia2021local,jia2022empty, guo2022effect}. They have uncovered reinforcement learning facilitates an optimal interaction intensity for cooperation~\cite{song2022reinforcement}, and Levy noise~\cite{wang2022levy}, reputation-based interaction~\cite{RZ:2023}, experienced guiders~\cite{YYWZCZ:2023} as well as local and global stimuli are able to promote cooperation in the PDGs with reinforcement learning~\cite{jia2021local}.


In spite of the progresses in the employment of reinforcement learning in explaining how humans deal with various
tasks~\cite{tomov2021multi}, there are still a number of interesting questions about the cooperation mechanism:
\emph{Can we develop a framework for reinforcement learning repeated games between two agents as a basis for extending it to a population? Can the results of two-agent reinforcement learning repeated games provide some useful insights for understanding cooperation in our society?}
Addressing these questions is of paramount significance because it helps us understand the emergence and maintenance of cooperation in reality from the perspective of artificial intelligence.

This paper is organized as follows. In Sec.~\ref{sec:model}, we present a general model that combines reinforcement learning algorithms with repeated games for two agents. The simulation results of strict prisoners' dilemmas game are shown in Sec.~\ref{sec:results}. To investigate the mechanism of the phenomena, we make some analysis in Sec.~\ref{sec:analysis}, which consists of four parts.  Finally, the conclusions and discussion are given in Sec.~\ref{sec:conclusions}.

\section{Reinforcement Learning for Repeated Games} \label{sec:model}

We start by introducing a general Reinforcement Learning Repeated Game (RLRG) for two agents, say ``Iris'' and ``Jerry'' (abbreviated as ``$i$'' and ``$j$'' thereafter), specifically they adopt the Q-learning algorithm~\cite{WD:1992}.
In this algorithm, Iris/Jerry may take an action against its co-player from an action set $\mathcal{A} = \set{a_{1}, \cdots, a_{n_a}}$ where it is in one of $n_s$ states from the state set $\mathcal{S} = \set{s_{1},\cdots,s_{n_s}}$. The goal is to find a policy that maximizes the expected cumulative reward. 
At $\tau$th round, the state of each agent consists of its own and its co-player's actions in the previous round, i.e. $s(\tau) = a(\tau-1)\tilde{a}(\tau-1)$, where $a$ and $\tilde{a}$ denote the agent and its co-player's actions, respectively. In other words, the state is the combination of the agent's and opponent's previous actions. Therefore, the state set is the Cartesian product of action set $\mathcal{A}\times\mathcal{A}\rightarrow \mathcal{S}$. 

In the Q-learning algorithm~\cite{WD:1992}, Iris/Jerry seeks for optimal policies through the so-called Q-table by learning. Here, the Q-table is a matrix on Cartesian product for states (columns) – actions (rows) $\mathcal{S}\times \mathcal{A}\rightarrow \mathbb{R}$:
\begin{eqnarray}
	{\bf Q}(\tau) =
	\left(                 
	\begin{array}{ccc}   
		Q_{s_{1},a_{1}}(\tau) & \cdots & Q_{s_{1},a_{n_a}}(\tau)\\  
		\vdots       & \ddots & \vdots   \\
		Q_{s_{n_s},a_{1}}(\tau) & \cdots & Q_{s_{n_s},a_{n_a}}(\tau)\\  
	\end{array}
	\right). 	\nonumber
\end{eqnarray}
With a Q-table in hand, Iris/Jerry takes action following its own Q-table 
\begin{eqnarray}
	a(\tau) \rightarrow \arg \max\limits_{a'}\{Q_{s,a'}(\tau)\}, a'\in\mathcal{A},  \label{eq:action}
\end{eqnarray}
with probability $1-\epsilon$, or a random action within $\mathcal{A}$ otherwise. Here, $\arg\max\limits_{a'}\{Q_{s,a'}(\tau)\}$
is the action corresponding to the maximum Q-value in the row of state $s$. The parameter $0 < \epsilon \ll 1$ is to introduce some random exploration besides the exploitation of the Q-table.

When Iris and Jerry make their decisions, they receive their own rewards according to their actions and a payoff matrix defined by the game they are playing
\begin{eqnarray}
	{\bf \Pi} =
	\left(                 
	\begin{array}{ccc}   
		\Pi_{a_{1}a_{1}} & \cdots & \Pi_{a_{1}a_{n_a}}\\  
		\vdots       & \ddots & \vdots   \\
		\Pi_{a_{n_a}a_{1}} & \cdots & \Pi_{a_{n_a}a_{n_a}}\\  
	\end{array}
	\right), \nonumber
\end{eqnarray}
where $\Pi_{a{\tilde a}} = \Pi(\tau)$ denotes the agent's reward if the agent with action $a$ is against action $\tilde{a}$ of its co-player.

At the end of $\tau$th round, Iris/Jerry update the element $Q_{s,a}$ for its $Q$-table  as follows
\begin{eqnarray}
	Q_{s,a}(\tau + 1) &= & g({\bf Q}(\tau),r(\tau)) \nonumber\\
	&=&(1-\alpha)Q_{s,a}(\tau) +\alpha\left[\gamma Q_{s',a'}^{\max}(\tau)+r(\tau)\right], \label{eq:update_Q}
\end{eqnarray}
where $\alpha \in (0, 1]$ is the learning rate reflecting the strength of memory effect, a large value of $\alpha$ means that the agent is forgetful since its previous value of $Q_{s,a}(\tau)$ is quickly modified. $r(\tau) = \Pi(\tau)=\Pi_{a(\tau)\tilde{a}(\tau)}$ is the agent's reward received in the current round $\tau$. 
$\gamma \in [0,1)$ is the discount factor determining the importance of future rewards since $Q_{s',a'}^{\max}$ is the maximum element in the row of next state $s'$ that could be expected. 
In such a way, the Q-table is updated, and the new state becomes $s(\tau+1) = a(\tau)\tilde{a}(\tau)$, and a single round is then completed.

To summarize, the pseudo code is provided in Algorithm~\ref{algorithm:protocol}.
\begin{algorithm}[htbp!]  
	\caption{The reinforcement learning for repeated games}\label{algorithm:protocol}
	\LinesNumbered 
	\KwIn{$\alpha$, $\gamma$, $\epsilon$ and game parameters}
	Initialization\;
	\For{Iris (i) and Jerry (j)}{
			Pick an action randomly from $\mathcal{A}$\;
			Create a Q-table with each item in the matrix near zero\;
	}
	\For{Iris (i) and Jerry (j)}{
	Generate state:$s \rightarrow a\tilde{a}$\;
	}
	\Repeat{\text the system becomes statistically stable or evolves for the desired time duration}
	{\For{Iris (i) and Jerry (j)}{
			Generate a random number $p$\;
			\eIf{$p<\epsilon$}{
				Pick an action randomly from $\mathcal{A}$
			}{
				Take an action according to state and Eq.~(\ref{eq:action})
			}
		}
		\For{Iris (i) and Jerry (j)}{
			Get reward according to their actions and payoff matrix\;
			Update Q-table according to Eq.~(\ref{eq:update_Q}) and state $s \rightarrow a\tilde{a}$\;
		}
	}
\end{algorithm}

\section{Simulation Results for Prisoner's Dilemma Game}\label{sec:results}
In this work, Iris and Jerry play the Strict Prisoner's Dilemma Game (SPDG) within our RLRGs framework, and we focus on the evolution of the cooperation preference $f_{c}$. Specifically, the action and state sets are respectively $\mathcal{A} = \set{a_{1},a_{2}}=\set{C, D}$, $\mathcal{S} = \set{s_{1},s_{2},s_{3},s_{4}}=\set{CC, CD, DC, DD}$, and correspondingly the time-evolving Q-table each player 
	\begin{eqnarray} 
		{\bf Q}(\tau) =
		\left(                 
		\begin{array}{cc}   
			Q_{s_{1},a_{1}}(\tau) & Q_{s_{1},a_{2}}(\tau)\\  
			Q_{s_{2},a_{1}}(\tau) & Q_{s_{2},a_{2}}(\tau)\\
			Q_{s_{3},a_{1}}(\tau) & Q_{s_{3},a_{2}}(\tau)\\
			Q_{s_{4},a_{1}}(\tau) & Q_{s_{4},a_{2}}(\tau)\\
		\end{array}
		\right)
		= 
		\left(                 
		\begin{array}{cc}   
			Q_{cc,c}(\tau) & Q_{cc,d}(\tau)\\  
			Q_{cd,c}(\tau) & Q_{cd,d}(\tau)\\
			Q_{dc,c}(\tau) & Q_{dc,d}(\tau)\\
			Q_{dd,c}(\tau) & Q_{dd,d}(\tau)\\
		\end{array}
		\right).
		\label{eq:Q_table}
	\end{eqnarray}
For SPDG, the payoff matrix in the Sec.~\ref{sec:model} is rewritten as 
\begin{eqnarray} \label{eq:payoff_matrix}
	{\bf \Pi} =
	\left(
	\begin{array}{cc}
		\Pi_{cc} & \Pi_{cd}\\
		\Pi_{dc} & \Pi_{dd}\\
	\end{array}
	\right)
	=
	\left(
	\begin{array}{cc}
		R & S\\
		T & P \\
	\end{array}
	\right)
	=
	\left(
	\begin{array}{cc}
		1& -b \\
		1+b & 0 \\
	\end{array}
	\right), 
\end{eqnarray}
where ${\bf \Pi}$ is with a tunable game parameter $b\in (0, 1)$, controlling the strength of the dilemma. A larger value of $b$ means a higher temptation to defect where cooperators are harder to survive.

Here, we define the average cooperation preference for Iris and Jerry within $t$th window in the simulation as follows 
\begin{eqnarray}\label{eq:ave_pre}
	f_{c}(t):=\frac{\sum\limits_{\tau=t-T}^{t}\sum\limits_{k\in\{i,j\}}\delta \left(a^{k}(\tau) - C\right)}{2T},
\end{eqnarray}
where $\delta(\cdots)$ is the Dirac delta function and $a^{i,j}$ are Iris or Jerry's actions. As can be seen that a sliding window with the length of $T$ is used for averaging. 
The time series of average preference can help us for better monitoring the evolution trend of different actions.
As $t$ and $T$ tend to infinity, $f_{c}(t)$ is the average cooperation preference over all time and can be denoted as $\bar{f}_{c}$. In our practice, we use sufficiently large $t$ and $T$ instead of infinity.

Apart from the average cooperation preference, we are also interested in the degree of fairness for the two agents.  For example, in the case of $C$-$D$ pair, the defector takes advantage over the cooperator, yielding an unfair reward division. To measure the degree of \emph{unfairness}, we defined it as the average reward difference between Iris and Jerry within consecutive rounds
\begin{eqnarray}\label{eq:Delta_r}
	\overline{\Delta R} := \frac{\sum\limits_{\tau=t-T}^{t}\bigg|\sum\limits_{\tau^{\prime}=\tau-1}^{\tau}
		\Pi^{i}(\tau^{\prime}) 
		- \Pi^{j}(\tau^{\prime})\bigg|}{T},
\end{eqnarray}
in which $\Pi^{i,j}$ are the rewards for Iris and Jerry. Obviously, when $\overline{\Delta R}\rightarrow 0$, it means the two agents statistically keep the action symmetry with each other, without apparent exploitation detected. Otherwise, a symmetry-breaking in their action/reward is present. In other words, the presence of unfairness means the symmetry-breaking of actions.

\begin{figure}[htbp!]
	\centering
	\includegraphics[width=1.0\textwidth]{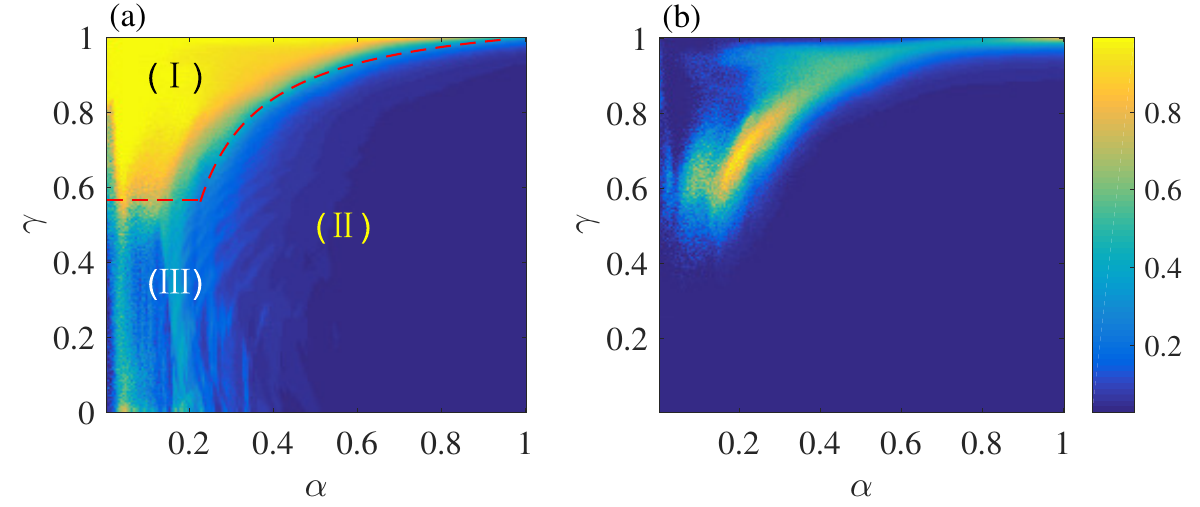}
	\caption{{\bf Averaged cooperation preference and averaged reward gap between agents over runs in the space of learning parameters.} 
	In (a), the cooperation preference for Iris and Jerry divides the parameter space of $\tuple{\alpha, \gamma}$ into three regions: high cooperation (\RNum{1}), full defection (\RNum{2}) and low cooperation (\RNum{3}). The boundaries of Region \RNum{1} is given by the theoretical prediction of Eq.~(\ref{eq:constraint_gamma}) and (\ref{eq:constraint_alpha_gamma}), see red dashed lines. In (b), the reward difference between agents, $\langle\overline{\Delta R}\rangle$, is only visible around the boundaries between Regions \RNum{1}--\RNum{2}, and Regions \RNum{1}--\RNum{3}.
	In (a-b), the shared parameters
	are $b = 0.2$ and $\epsilon = 0.01$. $\langle \bar{f}_{c}\rangle$ and $\langle\overline{\Delta R}\rangle$ is an average of $50$ runs from different initializations. 
	The scale and location of the window are $T = 6\times 10^6$ and $t = 8\times 10^6$ 
	after discarding the transient state including $4\times 10^6$ rounds.
	In the simulation, we divide the transient state into two stages and set $\epsilon=0.1$ and 
	$0.01$ in the first and second stages.} 
\label{fig:phase_diagram}  
\end{figure}

By fixing the dilemma strength $b = 0.2$, we firstly provide the average cooperation preference $\langle\bar{f}_{c}\rangle$ in the parameter domain of $(\alpha,\gamma)$ in Fig.~\ref{fig:phase_diagram}(a). The results show the domain can be roughly divided into three regions. In Region \RNum{1}, where $\alpha$ is low and $\gamma$ is high, the two agents maintain a high level of cooperation, showing that a strong memory effect and the long-sight facilitate the cooperation to thrive. By contrast, the opposite setup where agents are both forgetful and short-sighted results in a low cooperation preference, full defection is seen in Region \RNum{2}. Starting from Region \RNum{1}, $\langle\bar{f}_{c}\rangle$ decreases as the agents gradually become short-sighted (i.e. by decreasing $\gamma$), but cooperation does not disappear as long as the value of $\alpha$ is low enough, which is Region \RNum{3}. This means that cooperators still survive as long as the agents are not forgetful. In addition, 
Fig.~\ref{fig:phase_diagram}(b) also provides the average reward difference $\langle\overline{\Delta R}\rangle$ in the same domain as in Fig.~\ref{fig:phase_diagram}(a). 
We can see that the reward difference is almost zero within all the three Regions (I, II, and III), except at the boundaries between Regions \RNum{1} -- \RNum{2}, and Regions \RNum{2} -- \RNum{3}.
This means high unfairness (corresponding to frequent appearance of $C$-$D$ or $D$-$C$
cases) is observable only in the domain close to the boundary
between \RNum{1} and \RNum{2}; otherwise fairness is well maintained.

\begin{figure*}[htbp!]
	\centering
	\includegraphics[width=0.9\textwidth]{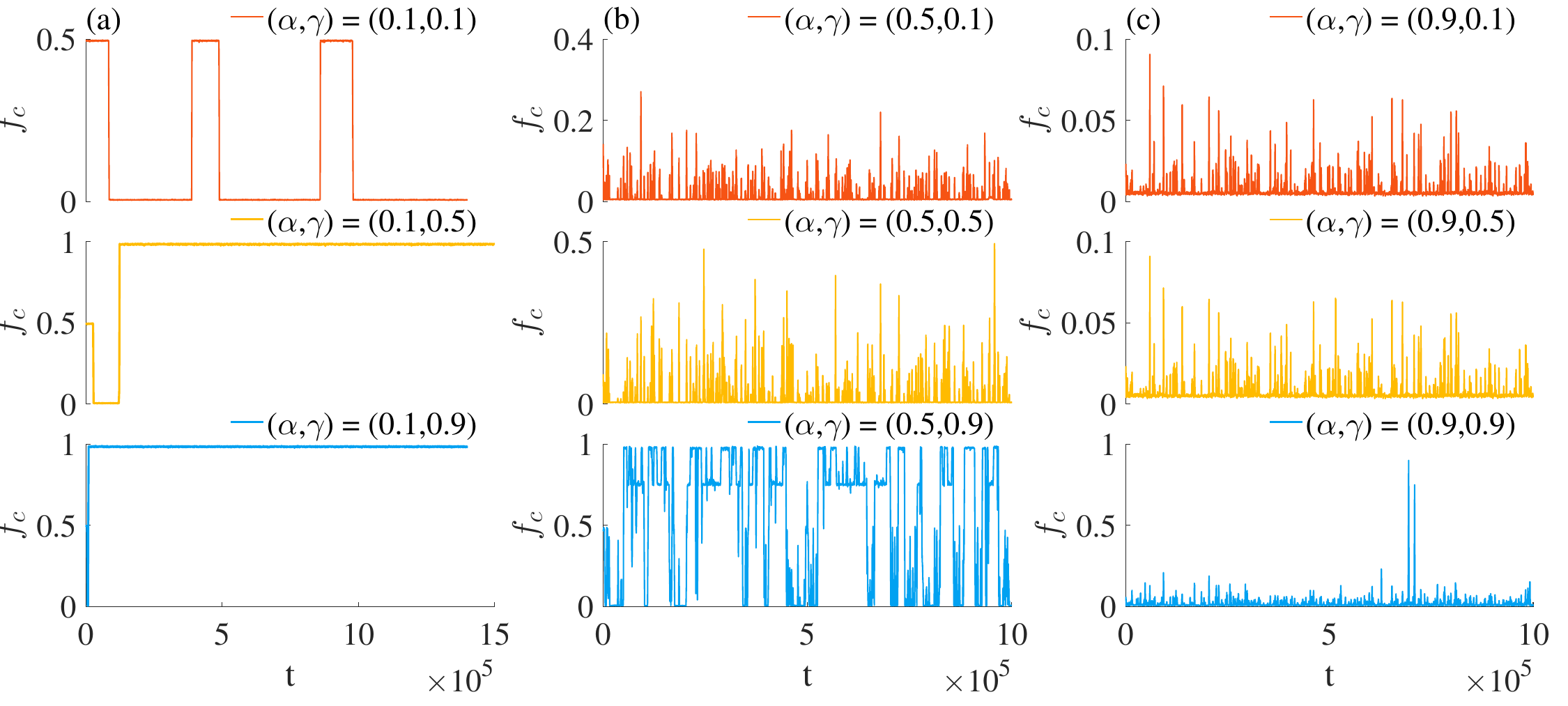}
	\caption{{\bf Typical time series of cooperation preference $f_{c}$ for nine different combinations of learning parameters.} In each row, the learning rates $\alpha = 0.1, 0.5, 0.9$ are increasing from left to right, but $\gamma$ is fixed, while for each column, the discounting factor $\gamma = 0.1, 0.5, 0.9$ are increasing from top to bottom but $\alpha$ are remain the same. The results show that the average cooperation preference $\bar{f}_{c}$ and the volatility of $f_{c}(t)$ are increased with $\gamma$, but decreased with $\alpha$. In (a-c), other parameters: $\epsilon = 0.01$, $b = 0.2$ and the length of time windows $T = 10000$.
	} 
	\label{fig:time_series} 
\end{figure*}

Fig.~\ref{fig:time_series} first shows some typical time series of cooperation preference for different combinations of $(\alpha, \gamma)$ in Fig.~\ref{fig:phase_diagram}(a). As can be seen, for a strong memory (small $\alpha$), the cooperation preference is relatively stable and $f_{c}$ increases with the expectation of future $\gamma$ [Fig.~\ref{fig:time_series}(a)]. With the increase of $\alpha$, a significant decrease in $f_{c}$ is detected, and the preference becomes quite volatile [Fig.~\ref{fig:time_series}(b)].  As $\alpha$ continues to increase, the cooperation is almost completely unsustainable in all three cases in [Fig.~\ref{fig:time_series}(c)]. 
By comparison, a high value of $\gamma$ is more likely to yield a high level of cooperation for a fixed learning rate.
These results and Fig.~\ref{fig:phase_diagram}(a) indicate that the combination of a relatively strong memory (a small learning rate $\alpha$) and a long vision for the future (a large discount factor $\gamma$) is the ideal scenario to sustain the cooperation.

\section{Mechanism Analysis}\label{sec:analysis}
\subsection{Coordinated optimal policies and modes}\label{subsec:analysis}

In the classical Q-learning, agent optimizes the value function of state-action pairs for the optimal policy $\pi^{*}$ to maximize the total expected cumulative reward. The value function (i.e. the Q-table), is characterized by a set of Bellman equations as
\begin{eqnarray}
Q^{\pi^*}_{sa} &=& \sum_{s^{\prime}\in\mathcal{S}}p(s'|s,a)\left[r(s,a,s') +\gamma\sum_{a^{\prime}\in\mathcal{A}}{\pi^{*}}(s^{\prime},a^{\prime})Q^{\pi^*}_{s^{\prime},a^{\prime}}\right]
\label{eq:Bellman_equations}
\end{eqnarray}
Here, $p(s'|s, a)$ is the transition probability for the agent from state $s$ to $s'$ when it takes action $a$ at $s$, $r(s,a,s')$ is the reward received for the agent. In a static environment, e.g. walking the labyrinth game, the optimal policy $\pi^*$ found by the agent is fixed because of the fixed environment of labyrinth thus also the time-independence of $p(s'|s,a)$. 

But, this is not the case in RLRGs. Because the environment is now composed of agents' states, which are time-varying. It's thus obvious that $p(s'|s,a)$ for Iris/Jerry is time-dependent and co-evolves with the Q-tables of both agents together with their policies~\cite{zhang2020understanding}.
This means that Iris and Jerry must seek for their optimal policy in a coordinated way with the other.
Here we define a set of {\em Coordinated Optimal Policies} (COP) denoted as $\tuple{\pi^{i*},\pi^{j*}}$, where $\pi^{i*}$ is optimal for Iris only if Jerry's policy is $\pi^{j*}$, and vice versa. For a COP, $\pi^{i,j*}$ is a set containing the optimal actions for Iris/Jerry at different state for the corresponding learning parameters.

\begin{figure*}[htbp!]
\centering
\includegraphics[width=0.9\textwidth]{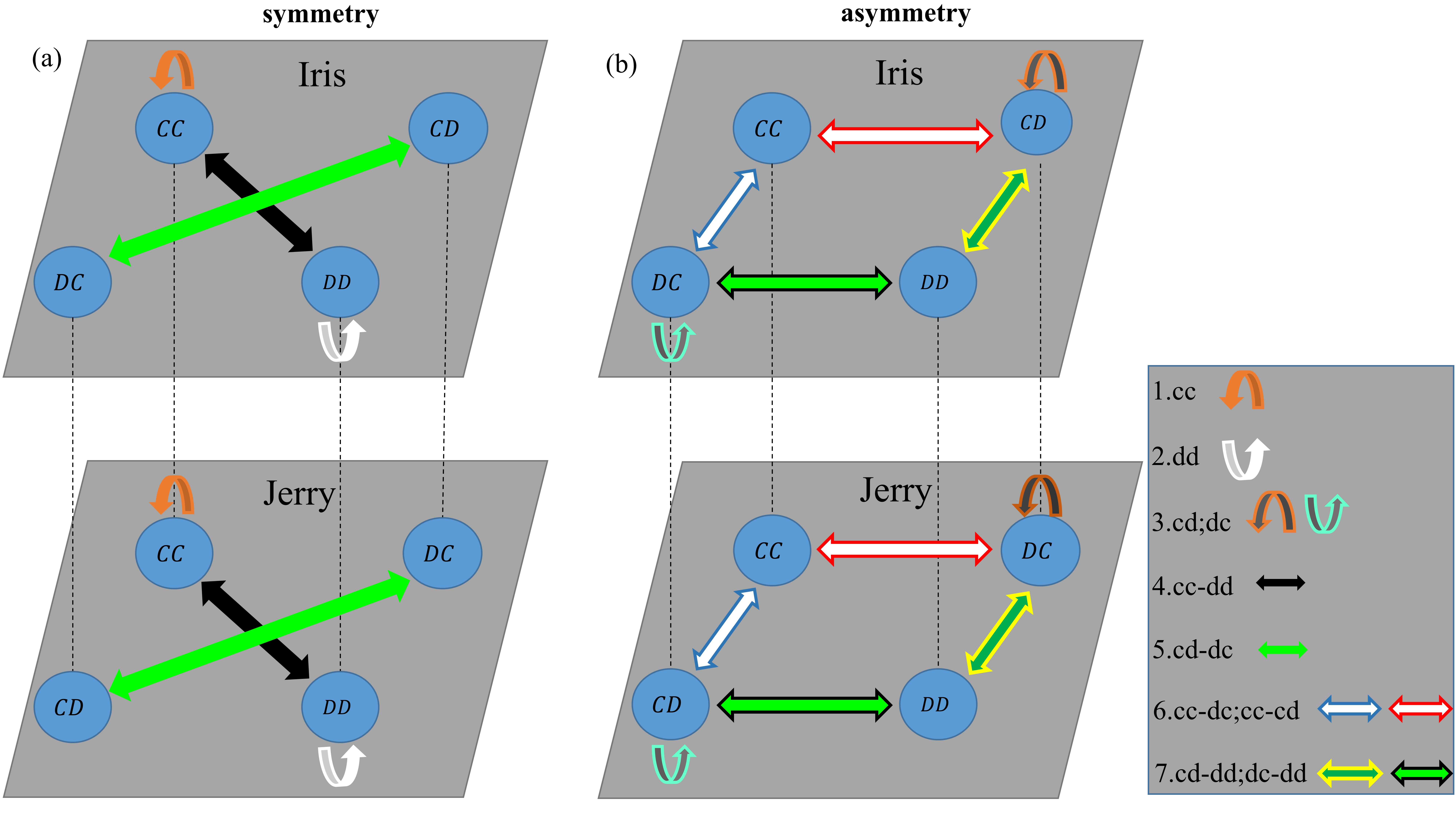}
\caption{{\bf Seven simple modes of state transitions for the strict prisoner's dilemma game.}  
	(a) Four symmetric modes include $m_1$: $\tuple{CC, CC}$, $m_2$: $\tuple{DD, DD}$, $m_3$: $\tuple{CC\leftrightarrow DD, CC\leftrightarrow DD}$ and $m_4$: $\tuple{CD\leftrightarrow DC, DC\leftrightarrow CD}$. 
	(b) Three asymmetric modes that are $m_5$: $\tuple{CD, DC}$, $m_6$: $\tuple{CC\leftrightarrow DC, CC\leftrightarrow CD}$ and $m_7$: $\tuple{DD\leftrightarrow CD, DD\leftrightarrow DC}$. 
	In the evolution, the modes evolve circularly between states shown in the upper and bottom rows.
}  
\label{fig:transition_modes}  
\end{figure*}

The COPs are able to remain unchanged for some time, then the system will fall into some corresponding {\em Coordinated Optimal Modes} (COMs), which consist of circular state transitions. Here, we check the COMs instead of directly examining the COPs.  By careful examination, we find that the system falls into a few modes once Iris/Jerry's policy $\pi^{i,j}$ remain unchanged for some time, which can be classified into $12$ circular modes as $\epsilon\rightarrow 0$ [see Table~\ref{tab:co_modes} in \ref{subsec:ap_convergence_modes}]. Fig.~\ref{fig:transition_modes} gives seven short ones from $m_1$ to $m_7$. Note that, due to the presence of exploration, long modes generally are more unstable compare to the shorter ones [see \ref{subsec:ap_stableQ}].

Here these modes can be classified according to their state symmetry. A mode is called symmetric if the state experience is statistically the same if Iris and Jerry are interchanged, otherwise it is called asymmetric.  For example, the mode of $\tuple{CC,CC}$ or $\tuple{DD, DD}$ in Fig.~\ref{fig:transition_modes}(a) is obviously symmetric, $\tuple{CD \leftrightarrow DC, DC \leftrightarrow CD}$ is also in symmetry.
But $\tuple{CC \leftrightarrow CD, CC \leftrightarrow DC}$ is asymmetric, since one agent always acts as a cooperator while the other adopts between cooperation or defection periodically. 
In an asymmetric mode, there is an unfairness $\Delta R$ between Iris and Jerry, and this unfairness is disappeared in symmetric modes. Note that, due to the presence of transition between two states, we need to compute the accumulated rewards for two consecutive rounds as the definition given in Eq.~(\ref{eq:Delta_r}). 
Here, COMs can be analogous to the equilibriums in the finite classical repeated PDGs~\cite{luce1989games,roth1978equilibrium,murnighan1983expecting}. According to observation, we learn that the number of COMs lies between $1$ and $4$ for a COP. 


\subsection{Transition of States}\label{subsec:state_transition}

\begin{figure}[htbp!]
	\centering
	\includegraphics[width=0.75\textwidth]{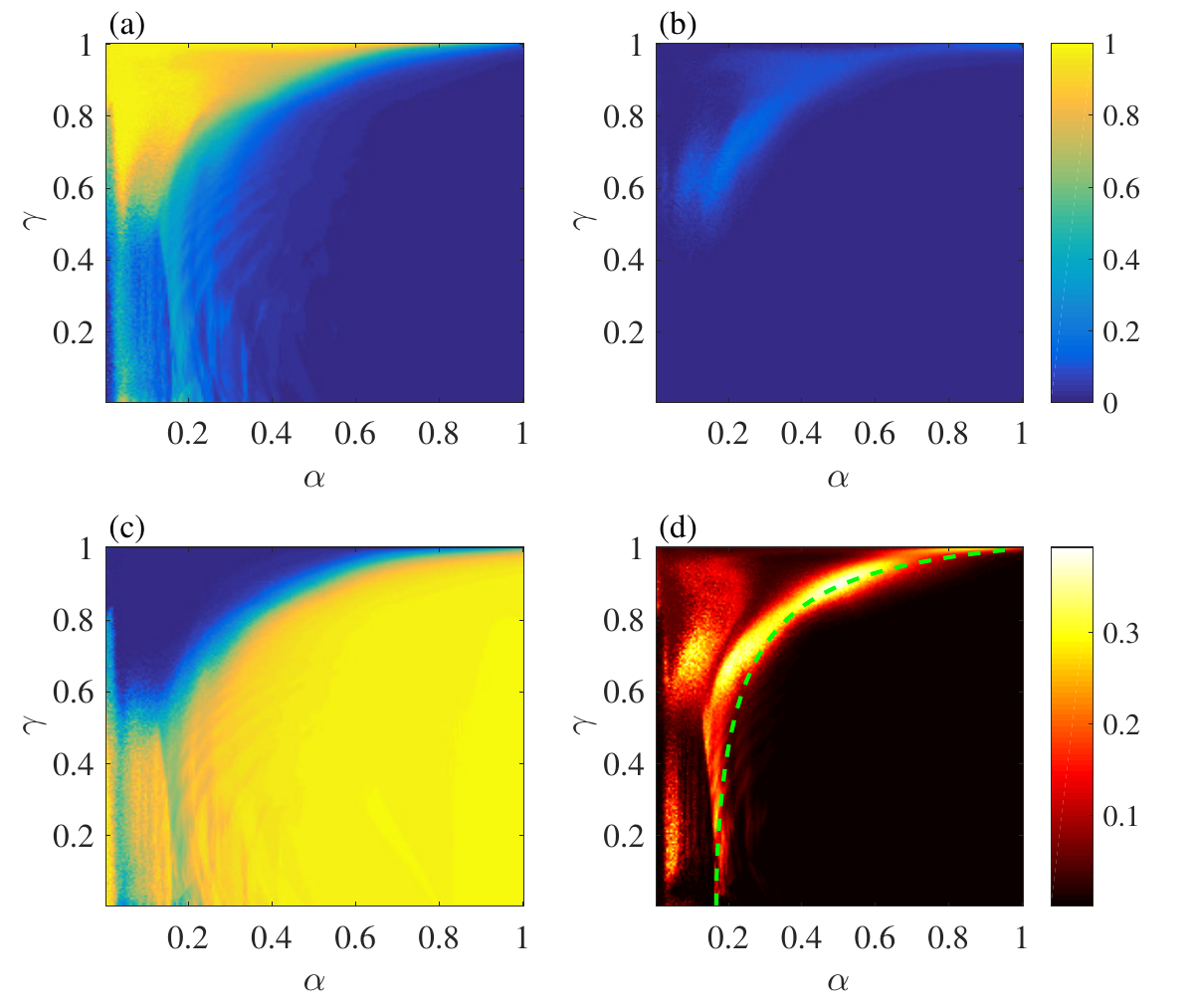}
	\caption{{\bf Probabilities of different states and mutual information between different consecutive states in the space of $\tuple{\alpha, \gamma}$.} (a-c) show the probabilities of states $p(CC)$, $p(CD)$ and $p(DD)$, and $p(DC)\approx p(CD)$, which is thus not shown. The results demonstrate that $CC$ and $DD$ are the dominant states in Regions \RNum{1} and \RNum{2} of Fig.~\ref{fig:phase_diagram}, while $CD$ and $DC$ only emerge around the boundary between \RNum{1} and \RNum{2}. The mutual information between different consecutive states are shown in (d), where the value is significantly enhanced at the boundaries of different regions, guided by the green dashed line.	
	The simulation details and parameters are exactly the same as in Fig.~\ref{fig:phase_diagram}
	} 
	\label{fig:state_distribution}  
\end{figure}

The above analysis indicates that the probability of state, $p(s)$, and the probabilities of state transitions, $p(s^{\prime}|s)$, reflects agents' COPs and corresponding COMs. Since the states and state transitions
for both agents can be mapped to each other as updating protocol shown, we only need to examine the cooperation preference of one agent through its $p(s)$ and $p(s^{\prime}|s)$. In here,  
we show the distribution of states in the parameter space of $\tuple{\alpha,\gamma}$ in Fig.~\ref{fig:state_distribution} (a-c) firstly. 
The state of $CC$ and $DD$ dominate in Region~\RNum{1} and~ \RNum{2} respectively, while the two coexist in Region~\RNum{3}. However, $CD$ and $DC$ appear only at the boundary between \RNum{1} and \RNum{2}. 

To characterize the correlation between the consecutive states, we compute the mutual information $I(s;s^{\prime})$ defined as 
\begin{eqnarray}
	I(s;s^{\prime}):=\sum\limits_{s\in\mathcal{S}}\sum\limits_{s^{\prime}\in\mathcal{S}}p(s,s^{\prime})\log \frac{p(s,s^{\prime})}{p(s)p(s^{\prime})},
	\label{eq:mutual_information}
\end{eqnarray}
which are shown in Fig.~\ref{fig:state_distribution} (d).
Results show that the mutual information between consecutive states is weak in the Region~\RNum{1} or~\RNum{2}, but is strong at their boundary. Strong mutual information implies that strong correlation between consecutive states and thus predictability in time.
It is well known that the dynamics near the criticality has long-term correlations, and a tiny perturbation is able to trigger a series of large fluctuations. Thus, the observations suggest that there is a bifurcation at the boundary in $\tuple{\alpha, \gamma}$, where the COP and the cooperation preference gradually changes as the learning parameters are varied.  

\begin{figure}[htbp!]
\centering
\includegraphics[width=0.75\textwidth]{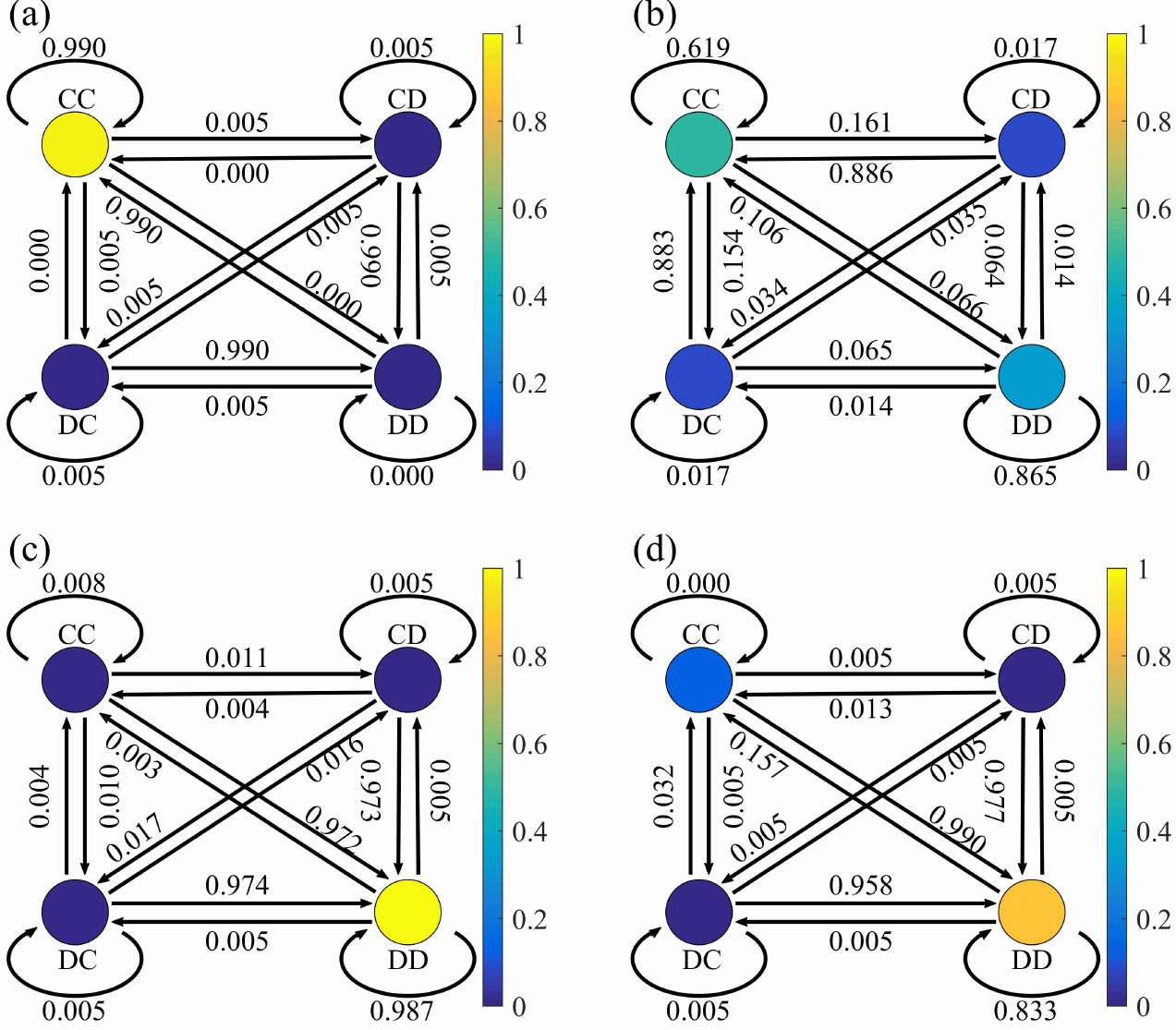} 
\caption{{\bf The probabilities of state and transition probabilities among the four states for four different parameter combinations.}		
The learning parameter combinations are $(\alpha,\gamma) = (0.1, 0.9)$, $(0.5, 0.9)$, $(0.9, 0.5)$ and $(0.1, 0.1)$, respectively in (a-d). 
The state probabilities $p(s)$ are encoded with colors, and the transition probabilities $p(s'|s)$ are explicitly given along the transition arrows. The four subplots are respectively located in Region \RNum{1}, the boundary between \RNum{1}--\RNum{2}, Regions \RNum{2} and \RNum{3} within Fig.~\ref{fig:phase_diagram}.	   
Other parameters: $\epsilon = 0.01$ and $b = 0.2$. } 
\label{fig:state_transition}
\end{figure}

Fig.~\ref{fig:state_transition} exhibits $p(s)$ and $p(s'|s)$ of one agent for four typical combinations of $(\alpha, \gamma)$ after the system becomes statistically stable. Specifically, we choose $\tuple{\alpha, \gamma}$ from Region \RNum{1}, the boundary between Regions \RNum{1} and \RNum{2}, Region \RNum{2} and Region \RNum{3} in Fig.~\ref{fig:phase_diagram}, respectively. The observations are made respectively as following:

(\Rnum{1}) In Fig.~\ref{fig:state_transition} (a), where the parameters are located in Region~\RNum{1}, $CC$ is shown to be the primary state and the other states are rarely seen according to the probabilities of $p(s)$. This is because of the dominating transitions $CD\rightarrow DD\rightarrow CC$ and $DC\rightarrow DD\rightarrow CC$ are dominant according to  $p(s^{\prime}|s)$.
With these, one learns that $\pi^{i*}$ and $\pi^{j*}$ in the unique COP are the same for both that are ``win-stay, lose-shift'' (WSLS). It implies that, with the strong memory and high expectations of the future, the exploitation by the agent's defection incurs retaliation from its co-player, and that revenge will ultimately lower its payoff and the agent is forced to cooperate. 

(\Rnum{2}) For the case at the boundary between \RNum{1} and \RNum{2}, Fig.~\ref{fig:state_transition} (b) shows the state is mainly composed of $CC$ and $DD$ mixture, while the fractions of the rest states are negligible. The reason for the absence of $CD$ and $DD$ is due to the high probabilities of $CD\rightarrow CC$ and $DC\rightarrow CC$. This means that as memory weakens and the expectation of the future diminishes, the cooperator becomes more tolerant of the defection of his partner, but this tolerance also leads to more exploitation, i.e., the possibilities of $CC\rightarrow CD$ and $CC\rightarrow DC$ also increase compare to Fig.~\ref{fig:state_transition} (a). The observations indicate that, on one hand, agents use co-player's tolerance to get more rewards by defection, but on the other hand, they do not want to break cooperation completely. However, once both defect, they stay in defection with a large probability. Though, still there is some chance to rebuild cooperation by $DD\rightarrow CC$. The results suggest that tolerance is a precursor that cooperation become fragile.

(\Rnum{3}) When located in Region~\RNum{2}, Fig.~\ref{fig:state_transition} (c) shows the state $DD$ dominates. 
 The transitions to state $DD$ are also non-negligible, i.e. $CC\rightarrow DD$, $CD\rightarrow DD$, and $DC\rightarrow DD$. 
Consequently, agents
almost have no chance to rebuild cooperation once they have defected.
That is to say, both $\pi^{i*}$ and $\pi^{j*}$ are ``all-defection'' (All-D), i.e., the COP is $\tuple{\text{All-D}, \text{All-D}}$ in Region \RNum{2}, where both agents are of weak memory effect and low expectation of the future.

(\Rnum{4}) Fig.~\ref{fig:state_transition} (d) shows the scenario in Region~\RNum{3}, where the state is also mostly composed of $CC$ and $DD$ mixture, as in the case of Fig.~\ref{fig:state_transition} (b). However, compared to case ii, the $CC$ state now becomes unstable although the possibility to rebuild cooperation is non-zero. Overall, in this region, agents' policies preserve most features of the All-D, but also with the property of WSLS, due to the presence of $DD\rightarrow CC$.
Case (\Rnum{1}-\Rnum{3}) suggest a strong memory may be the prerequisites for rebuilding cooperation when the cooperation is broken.  

\subsection{Temporal Correlation}\label{subsec:co_modes}

To further capture the correlation in time between consecutive states, we compute the joint probability $p(s,s^{\prime})$ for the state transition from $s$ to $s^{\prime}$. And as the benchmark~\cite{hegland2007apriori}, we also compute the products of their state probabilities $p(s)p(s^{\prime})$. When $p(s,s^{\prime})>p(s)p(s^{\prime})$, it means that there is a positive correlation between the two consecutive states compare to the purely random occurrence, and vice versa.

\begin{figure}[htbp!]
\centering
\includegraphics[width=0.85\textwidth]{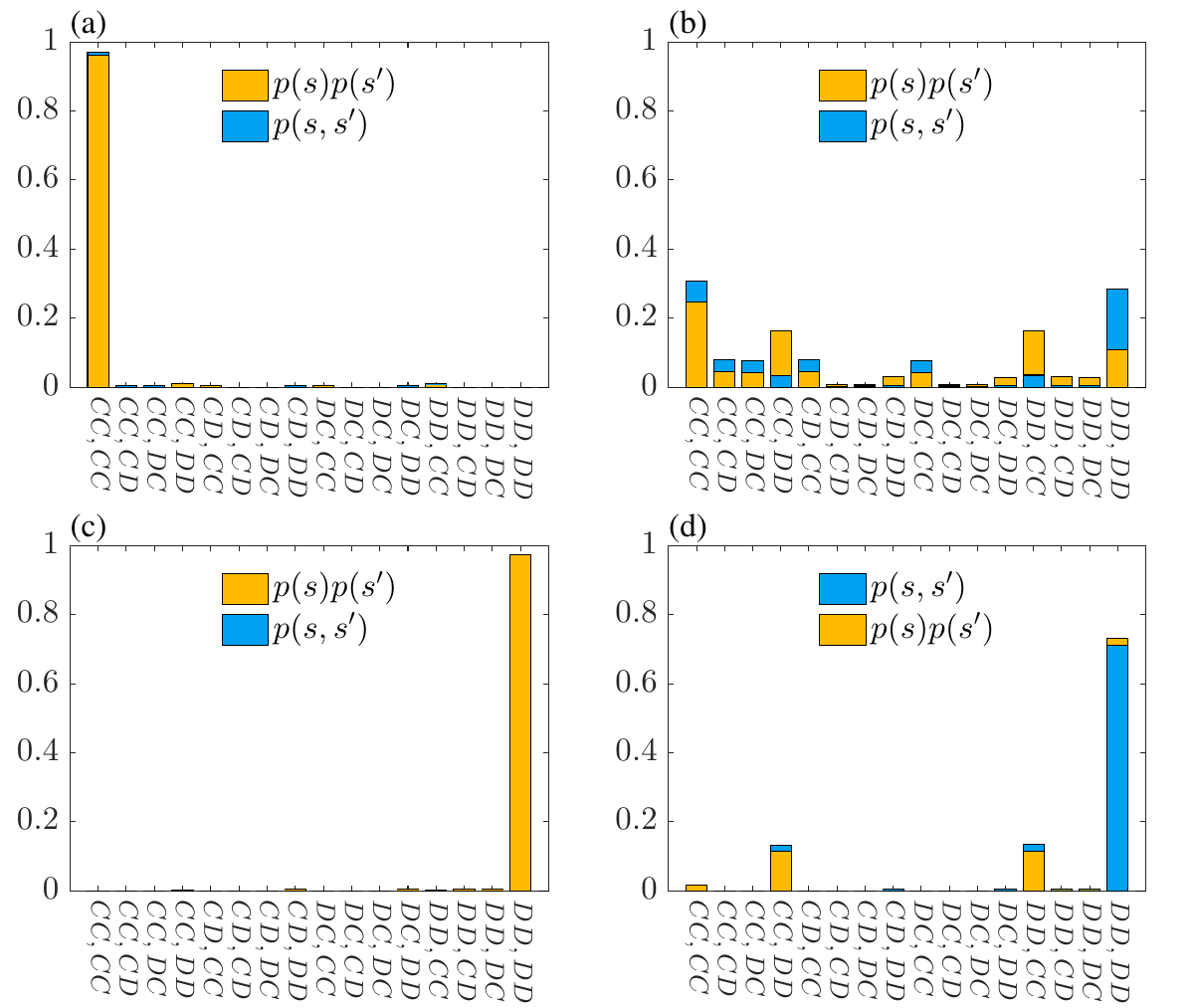}
\caption{{\bf The joint probability $p(s,s^{\prime})$ and the products of marginal probabilities between the consecutive states $p(s)p(s^{\prime})$ as the benchmark.} The parameters in (a-d) are exactly the same as in (a-d) of Fig.~\ref{fig:state_transition}. Here, the gap between $p(s,s^{\prime})$ and $p(s)p(s^{\prime})$ reflects temporal correlation between consecutive states $s$ and $s^{\prime}$. The four subplots are respectively located in Region \RNum{1}, the boundary between \RNum{1}--\RNum{2}, Regions \RNum{2} and \RNum{3} within Fig.~\ref{fig:phase_diagram}.	  	
The scale and location of the window are $T = 6\times 10^8$ and $t = 8\times 10^8$ after discarding the transient state including $2\times 10^8$ rounds.
} 
\label{fig:p_p}  
\end{figure} 

Figure.~\ref{fig:p_p} displays $p(s,s^{\prime})$, $p(s)p(s^{\prime})$ for the four typical cases with the same parameter combinations as in Fig.~\ref{fig:state_transition}. Figure.~\ref{fig:p_p}(a) and (c) show that $p\tuple{CC, CC}$ and $p\tuple{DD, DD}$ are the only dominant joint probabilities in Region~\RNum{1} and \RNum{2}, respectively. Since the gaps between $p(s,s^{\prime})$ and $p(s)p(s^{\prime})$ are almost invisible in these two cases, we compute their Cohen's kappa coefficients that are given in Fig.~\ref{fig:kappa} of the \ref{subsec:ap_kappa}, which verify positive correlations between consecutive $CC$s ($DD$s) in Region~\RNum{1} (\RNum{2}).
This observation means that the COMs of $\tuple{CC, CC}$ and $\tuple{DD, DD}$ are quite stable in Region~\RNum{1} and \RNum{2}, respectively. 

The reason why cooperation is preferred in Region~\RNum{1} is due to the fact that cooperation must build on the predictability of agents' policy towards each other; and the predictability relies on the stability of action preference which increases with $\gamma$ but decreases with $\alpha$ [Fig.~\ref{fig:time_series} and Eqs.~(\ref{eq:convergence_alpha}) and ~(\ref{eq:convergence_gamma})]. Therefore, a high level of cooperation is expected in Region~\RNum{1}, where agents are of strong memory effect (a small $\alpha$) and a long vision for the future (a large $\gamma$). By contrast, weakened cooperation is observable due to the inferior predictability to the opponent as the memory effect weakens ($\alpha$ increases) and the expectation of the future diminishes ($\gamma$ decreases).

At the boundary between Region~\RNum{1} and \RNum{2}, Fig.~\ref{fig:p_p} (b) shows that there are positive correlations between $CC$ and all states except $DD$, while $DD$ is only positively correlated with itself. This indicates that $\tuple{CC, CC}$ starts to become unstable and as the competing transition $\tuple{CC\leftrightarrow CD, CC\leftrightarrow DC}$ emerges, and $\tuple{DD, DD}$ starts to stabilize. The results imply that there is competition between tolerance and revenge for agents in dealing with the exploitation from its opponent at boundary.

In Region~\RNum{3}, Fig.~\ref{fig:p_p}(d) displays $p(DD,DD)$ is much higher than all other joint probabilities, the two states $CC$ and $DD$ coexist. While the correlations in $p(DD,DD)$ and $p(CC,CC)$ are both negative, the $p(CC,DD)$ and $p(DD,CC)$ are positive (see Fig.~\ref{fig:kappa} of \ref{subsec:ap_kappa} for their Cohen's kappa coefficients), which implies enhanced propensities of the transitions $\tuple{CC\leftrightarrow DD, CC\leftrightarrow DD}$ compared to the benchmark level.

Here we further analyze which factors influence the emergence of high cooperation in the Q-learning. It is well known that two-agent systems may not be able to achieve stable cooperation if they make Markovian decisions. However, we must clarify that although the update protocol of Q-learning is Markovian, each Q-element as a subspace of the whole Q-table is actually non-Markovian. The action decisions for both agents are thus non-Markovian, which is influenced by the memory length depending on learning parameters $\alpha$ and $\gamma$. This is the reason why a two-agent system is able to achieve stable cooperation under Q-learning for some learning parameters. In addition, the predictability of agents to opponents must be built on the history of the game and their memories. Thereby, the interaction history also plays a vital role in cooperation.

\subsection{Boundary of High Cooperation Level}\label{ap_subsec:analysis_PDGs}
Our simulation shows $\tuple{\text{WSLS},\text{WSLS}}$ and $\tuple{CC\leftrightarrow CD, CC\leftrightarrow DC}$ coexist at the boundary. According to the clue, we conjecture that there are two competitive balances at the boundary: (1) the selection between $\tuple{\text{WSLS},\text{WSLS}}$ and $\tuple{CC\leftrightarrow CD, CC\leftrightarrow DC}$; (2) the switch between $\tuple{CC\leftrightarrow CD, CC\leftrightarrow DC}$ and $\tuple{\text{WSLS},\text{WSLS}}$ with perturbations due to the exploration. 

For the first balance, it is the competition between the revenge of $\tuple{\text{WSLS},\text{WSLS}}$ and the tolerance of $\tuple{CC\leftrightarrow CD, CC\leftrightarrow DC}$ in dealing with exploitation. Thus, as the values for the revenge and the tolerance, $Q_{cd,d}^{w}$ and $Q_{cd,c}^{t}$ are our pivotal Q-values. The analysis in ~\ref{subsec:ap_stableQ} shows that the Q-values on a {\em key path} of COM/COP will converge to fixed values [as Eqs.~(\ref{eq:core_mode}) and ~(\ref{eq:core_policy}) shown]. Accordingly,  the converged Q-values for $\tuple{\text{WSLS}, \text{WSLS}}$
\begin{eqnarray}
	\left\{
	\begin{array}{l}
		Q_{cc,c}^{w*} = Q_{dd,c}^{w*} = \displaystyle{\frac{\Pi_{cc}}{1-\gamma}},\\
		Q_{cd,d}^{w*} = Q_{dc,d}^{w*} = \displaystyle{\Pi_{dd} + \gamma\frac{\Pi_{cc}}{1-\gamma}},
	\end{array}
	\right.
\end{eqnarray}
and for the mode $\tuple{CC\leftrightarrow CD, CC\leftrightarrow DC}$ are
\begin{eqnarray}
	\left\{
	\begin{array}{l}
		Q_{cc,c}^{t*} = \displaystyle{\frac{\Pi_{cd}+\gamma \Pi_{cc}}{1-\gamma^2}},\\
		Q_{cd,c}^{t*} = \displaystyle{\frac{\Pi_{cc}+\gamma \Pi_{cd}}{1-\gamma^2}},
	\end{array}
	\right. 
	\left\{
	\begin{array}{l}
		Q_{cc,d}^{e*} = \displaystyle{\frac{\Pi_{dc}+\gamma \Pi_{cc}}{1-\gamma^2}},\\
		Q_{dc,c}^{e*} = \displaystyle{\frac{\Pi_{cc}+\gamma \Pi_{dc}}{1-\gamma^2}}.
	\end{array}
	\right.
\end{eqnarray}
Due to the asymmetry of the COM, we use  superscript $e$ and $t$ to distinguish the exploiters and the tolerant agent's Q-values, respectively. At the boundary, the constraint for the tolerant agents is $Q_{cd,d}^{w*}=Q_{cd,c}^{t*}$, i.e.,    
\begin{eqnarray}
	\gamma = \frac{-1-b}{2}+\frac{1}{2}\sqrt{5+2b+b^2}. \label{eq:constraint_gamma}
\end{eqnarray}

\begin{figure}[htbp!]
	\centering
	\includegraphics[width=0.6\textwidth]{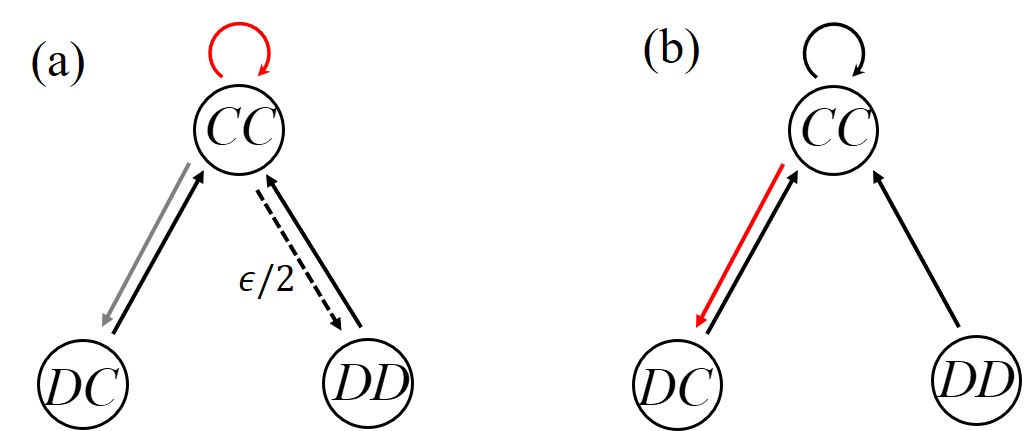}
	\caption{In (a), the diagram shows that the exploiter switches the state from $CC\rightarrow DC$ (gray arrow) in $\tuple{CC\leftrightarrow CD, CC\leftrightarrow DC}$ to $CC\rightarrow DD$ (dashed arrow) in $\tuple{\text{WSLS, WSLS}}$ after the tolerant agent defects by exploration. (b) demonstrates that cooperation is reestablished because the tolerant agent's exploration cause the mode switch from $\tuple{CC\leftrightarrow CD, CC\leftrightarrow DC}$ to $\tuple{CC, CC}$ of $\tuple{\text{WSLS, WSLS}}$. In the figure, the red and black arrows denote the blocked and unblocked transitions as $\epsilon \rightarrow 0$.} 
	\label{fig:mode_switch}  
\end{figure} 

For the second balance, after entering $\tuple{CC\leftrightarrow CD, CC\leftrightarrow DC}$, the exploiter confronts staying on the current COM or switching to the competing COM of $\tuple{\text{WSLS},\text{WSLS}}$ when faced with the tolerant agent's random exploration.
As a COM or COP, $\tuple{CC\leftrightarrow CD, CC\leftrightarrow DC}$ and $\tuple{\text{WSLS}, \text{WSLS}}$ obviously have some stability at the boundary. Thus, for 
the exploiter, the Q-values on the \emph{key paths} of $\tuple{CC\leftrightarrow CD, CC\leftrightarrow DC}$ and $\tuple{\text{WSLS},\text{WSLS}}$ have converged to the fixed values correspondingly before the random exploration, i.e., 
$Q_{cc,d} \rightarrow Q_{cc,d}^{e*}$, $Q_{dd,c} \rightarrow Q_{dd,c}^{w*}$ and  
$Q_{cc,c} \rightarrow Q_{cc,c}^{w*}$. If the tolerant agent defects at the state $CC$
under exploration, the balance of whether or not to change the path for the exploiter is decided by $Q_{cc,c}$ and $Q_{cc,d}$, i.e.,
\begin{eqnarray}
	Q_{cc,d} = (1-\alpha)Q_{cc,d}^{e*} + \alpha(\gamma Q_{dd,c}^{w*} + \Pi_{dd}) = Q_{cc,c}^{w*}
\end{eqnarray}
as Fig.~\ref{fig:mode_switch} shows.
So, another constraint for the boundary is $Q_{cc,d}=Q_{cc,c}^{w*}$, i.e., 
\begin{eqnarray}
	\alpha=\frac{b}{1+b-\gamma^2}. \label{eq:constraint_alpha_gamma}
\end{eqnarray}
After substituting $b = 0.2$, the temptation $b$ of Fig.~\ref{fig:phase_diagram}, into Eqs.~(\ref{eq:constraint_gamma}) and ~(\ref{eq:constraint_alpha_gamma}), the predicted boundaries well match the results of our simulations as shown Fig.~\ref{fig:phase_diagram}.

Our analysis shows that, with the increase of $\alpha$ and decrease
of $\gamma$, $\tuple{CC, CC}$ becomes unstable because tolerance replaces immediate revenge of WSLS in the face of exploitation from the opponent, leading to the degradation of cooperation. In addition, the analysis also confirms that a transition occurs at the boundary in Fig.~\ref{fig:state_distribution}(d), where $\tuple{\text{WSLS},\text{WSLS}}$ loses stability and COP changes gradually with the change of learning parameters.

\subsection{Impact of Random Exploration}\label{subsec:exploration}
\begin{figure}[htbp!]
	\centering
	\includegraphics[width=0.85\textwidth]{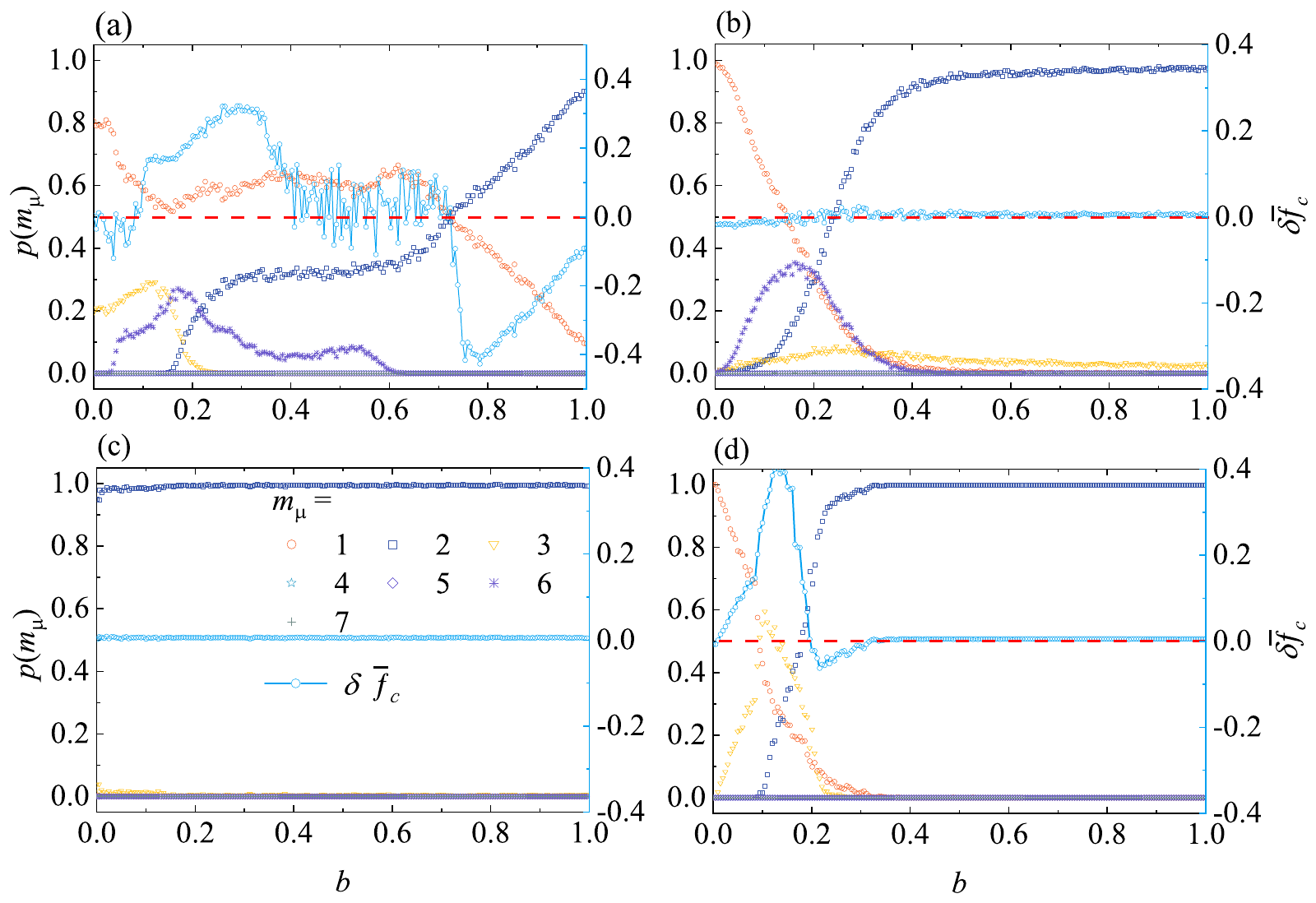}
	\caption{{\bf Probabilities of different modes and the average cooperation preference gap with and without exploration.} 
		The left $y$-axis shows the probabilities of $7$ short modes aforementioned in Fig.~\ref{fig:transition_modes} 
		that are $m_1$: $\tuple{CC, CC}$, $m_2$: $\tuple{DD, DD}$, $m_3$: $\tuple{CC\leftrightarrow DD, CC\leftrightarrow DD}$, $m_4$: $\tuple{CD\leftrightarrow DC, DC\leftrightarrow CD}$, $m_5$: $\tuple{CD, DC}$, $m_6$: $\tuple{CC\leftrightarrow DC, CC\leftrightarrow CD}$, and $m_7$: $\tuple{DD\leftrightarrow CD, DD\leftrightarrow DC}$. And the right $y$-axis
		shows the gap between the cooperation level with and without exploration, which is defined in Eq.~(\ref{eq:probabilities_mode}).  The red dashed line is $p(m_{\mu}) = 0.5$ or the gap $\delta \bar{f}_c = 0$ for comparison. As $\delta \bar{f}_c$ is above/below the line, this means that the exploration facilitates/suppresses the cooperation preference.
		The parameters in (a-d) are exactly the same as in (a-d) of Fig.~\ref{fig:state_transition}. In the simulations, the exploration stops after $4\times 10^6$ rounds and the initial exploration rate is $\epsilon = 0.01$. The probabilities of different modes in Eq.~(\ref{eq:probabilities_mode}) are obtained after $2000$ runs. 
	}\label{fig:probabilities_modes}  
\end{figure}

We further analyze the impact of the random exploration on cooperation. For comparison, we turn off the exploration (by setting $\epsilon=0$) after the evolution is stabilized to reveal the difference, which is shown as a function of the game parameter $b$, see Fig.~\ref{fig:probabilities_modes}.

When the exploration is ceased, the evolution fails to go through from one mode to another, but falls into a single mode.
As can be seen, only some of them e.g. $m_{1}$, $m_{2}$, $m_{3}$ and $m_6$ are present acting as COMs and the probabilities of the aforementioned short modes meet $\sum_{\mu=1}^{7}p(m_\mu) \approx 1$.
Fig.~\ref{fig:probabilities_modes}(a-b) and (d) show that $p(m_1)$ is decreasing as $b$ increases, while $p(m_2)$ is doing the opposite. $p(m_3)$ however shows a non-monotonic dependence on $b$ from increasing to decreasing. Furthermore, an asymmetric COM, $m_6$, appears in (a-b) but not in (c-d), and the mode also shows non-monotonic change. According to Fig.~\ref{fig:phase_diagram} (b), we conjecture that $m_6$, resulting from tolerance, is a metastable mode between $m_1$ and $m_2$. The reason is that (b) shows $m_6$ seemingly redundant around $b = 0.19$ as $\tuple{\alpha,\gamma} = \tuple{0.5, 0.9}$, which is at the boundary of \RNum{1} (see Fig.~\ref{fig:phase_diagram} and Eq.~(\ref{eq:constraint_alpha_gamma})). 

To investigate the impacts of exploration, we compute the
difference between the cooperation level with and without exploration 
\begin{eqnarray}
\delta \bar{f}_c &:=& \langle\bar{f}_{c}\rangle - \langle f_{c_0}\rangle \nonumber \\
&=& \langle\bar{f}_{c}\rangle-\sum_{\mu=1}^{7} p(m_\mu)f_{c}(m_{\mu}),
\label{eq:probabilities_mode}
\end{eqnarray}
where $\langle\bar{f}_{c}\rangle$ is the observed cooperation prevalence before the exploration is turned off, $f_{c}(m_{\mu})$ is cooperation preference of the corresponding mode $m_{\mu}$, e.g., the preferences for $m_ 3$ and $m_6$ are $f_c(m_3) = 0.75$ and $f_c(m_6) = 0.5$.
The second term as a benchmark is thus the expected cooperation prevalence in the absence of exploration. 

Fig.~\ref{fig:probabilities_modes} also plots $\delta \bar{f}_c$ as a function of temptation $b$ for the same parameter combination. Firstly, there is little impact of exploration in Fig.~\ref{fig:probabilities_modes}(b) and (c) since $\delta \bar{f}_c\approx 0$. However, this is not case in Fig.~\ref{fig:probabilities_modes}(a) and (d).
In both cases, the presence of exploration improves the cooperation preference at the small range of $b$. 
But, as $b$ increases, the exploration suppresses the cooperation preference in Fig.~\ref{fig:probabilities_modes}(a), but shows little impact in Fig.~\ref{fig:probabilities_modes}(d).
This is because exploration does not play a significant
role in the pure modes $m_1$ and $m_2$  as in (b, c). But
when multiple states coexist such as in (a-d) for an
intermediate $b$ value, the transition among the evolved states
yields a non-trivial impact of cooperation prevalence.
This is quite different from the previous work~\cite{barfuss2023intrinsic}, where the exploration always facilitates the cooperation under scheme of reinforcement learning.

\section{Conclusion and discussion}\label{sec:conclusions}
In the work, we introduce a general reinforcement learning for repeated dyadic games, where each agent optimizes it’s policies through Q-learning algorithms. Specifically, we focus on the impacts
of the learning rate and discount factor on the evolution of cooperation in the strict prisoner’s dilemma game.
We reveal that agents can achieve a high level of cooperation when they have a strong memory and a confident foresight for the future. However, cooperation is completely broken when the agents become forgetful or short-sighted. 

To proceed, we examine the agents' policies by checking their probabilities of states and states transitions. In the high cooperation region, both Q-tables exhibit WSLS property as their \emph{Coordinated Optimal Policy} (COP). 
In contrast, both agents are doomed
to defect when their COP is composed of All-D for both in defection region.
The most striking case occurs on the boundary of these two regions, where one agent tolerates its opponent’s defection and maintains cooperation, while the other
takes the advantage to maximize its own rewards. Such tolerance may be regarded as a precursor to the instability of cooperation.
A mixture of both WSLS-like and All-D-like policies
finds its niche when agents are endowed with a strong memory
but a short sight, which allows a low level of cooperation.

Moreover, analogous to the equilibriums of the finite repeated PDGs~\cite{luce1989games,roth1978equilibrium,murnighan1983expecting},  
we find that the agents' behavior can be decomposed into one of several circular \emph{Coordinated Optimal Modes} (COMs). The time correlation between consecutive states are also given, and the pronounced mutual information between consecutive states at the boundary indicates some sort of criticality relating to bifurcation of COPs.
Based on evolution of COMs and COPs, our theoretical analysis give the boundary of high cooperation and
verifies the indication by showing a decent match with the numerical results.
Finally, we also examine the effects of exploration rate on cooperation. In contrast to the previous work~\cite{barfuss2023intrinsic}, its impact depends on the composition of COMs, could be  positive, negative, or no influence at all.

In brief, by establishing an exploratory framework for the analysis of dynamics of RLRGs, we show some fundamentally interesting results. However, our findings leave too many questions unanswered. For example, an interesting perspective is to relate the COP to dynamical attractors, but a
proper formulation still needs to be shaped. Addressing this
question could help to obtain all COPs in more complex scenarios. 
What is more, our results uncover that COPs between two agents depend on their memory strengths and the
game history. This suggests that defection is the optimal policy for an agent with reinforcement learning against fixed no-memory strategies, such as All-Cooperation, All-Defection, and Random-Strategy, etc. Indeed, some work in the evolutionary game framework has also focused on the memory effect and found that longer-memory strategies would be a promising direction of direct reciprocity for cooperation~\cite{hilbe2017memory,murase2020five,li2022evolution}. However, it remains an open question whether similar direct reciprocity and cooperation emerge when the agent faces some classical memory strategies, such as win-stay-lose-shift, tit-for-tat, and zero-determinant strategies, etc. Furthermore, the researchers have extended selfless reciprocity, even cooperation, to moral behaviours and laid mathematical foundations for moral self preferences~\cite{capraro2021mathematical}. So, with the help of research based on reinforcement learning multi-agents and the foundations, can we navigate the complicated landscape of human morality in our society?
A further open question of special significance is to identify effective early-warning signals, where the theory of criticality may lend a hand to prevent irreversible and disruptive defective behaviours. 

\section*{Acknowledgments}
We are supported by the Natural Science Foundation of China under Grant No. 12165014 and the Key Research and Development Program of Ningxia Province in China under Grant No. 2021BEB04032. CL is supported by the Natural Science Foundation of China under Grant No. 12075144. 

\appendix

\counterwithin{figure}{section}
\section{More Details on COMs}

\subsection{Learning Parameters and Convergence}\label{subsec:ap_convergence_modes}

As stated in Subsec.~\ref{subsec:analysis}, the state transitions in our RLRG model will fall into one of circular modes as $\epsilon\rightarrow 0$ if Iris and Jerry's policies remain unchanged for some time. 
In Table.~\ref{tab:co_modes}, we list all modes of circular state transitions under SPDG setting, where ``cycle-$m$'' means the mode contains $m$-states, e.g., cycle-$1$ is a single-state self-loop. Besides, all modes in cycle-$3$ are asymmetric, while the mode in cycle-$4$ is symmetric. For cycle-$1$ and -$2$ modes,  the symmetric and asymmetric modes are separated by semicolons.
\begin{table}[htbp]
	\centering
	\caption{A list of modes in RPRG for the prisoner's dilemma.}\label{tab:co_modes}
	\resizebox{0.95\textwidth}{!}{\scriptsize\begin{tabular}{|cc|}
		\hline
		cycle-1 & \begin{tabular}[c]{@{}c@{}}  $\begin{pmatrix}
				CC,CC
			\end{pmatrix}$,
			$\begin{pmatrix}
				DD,DD
			\end{pmatrix}$; \\
			$\begin{pmatrix}
				CD,DC
			\end{pmatrix}$\end{tabular}                           \\ \hline
		cycle-2 & \begin{tabular}[c]{@{}c@{}}	$\begin{pmatrix}
				CC \leftrightarrow DD, CC \leftrightarrow DD
			\end{pmatrix}$,$ \begin{pmatrix}
				CD \leftrightarrow DC, DC \leftrightarrow CD
			\end{pmatrix} $; \\ 	$\begin{pmatrix}
				CC\leftrightarrow DC, CC \leftrightarrow CD
			\end{pmatrix}$,$ \begin{pmatrix}
				CD\leftrightarrow DD, DC \leftrightarrow DD
			\end{pmatrix}
			$\end{tabular}                           \\ \hline
		cycle-3 & \begin{tabular}[c]{@{}c@{}}$\begin{pmatrix}
				& CC             &                &                & CC             &                \\
				\qquad\nearrow &                & \searrow\qquad & \qquad\nearrow &                & \searrow\qquad \\
				CD             & \longleftarrow & DD,            & DC             & \longleftarrow & DD
			\end{pmatrix}$,\\$ \begin{pmatrix}
				& CC             &                &                & CC             &                \\
				\qquad\nearrow &                & \searrow\qquad & \qquad\nearrow &                & \searrow\qquad \\
				DD             & \longleftarrow & DC,            & DD             & \longleftarrow & CD
			\end{pmatrix} $,\\ $ \begin{pmatrix}
				& CC             &                &                & CC             &                \\
				\qquad\nearrow &                & \searrow\qquad & \qquad\nearrow &                & \searrow\qquad \\
				DC             & \longleftarrow & CD,            & CD             & \longleftarrow & DC
			\end{pmatrix} $,\\$\begin{pmatrix}
				& CD             &                &                & DC             &                \\
				\qquad\nearrow &                & \searrow\qquad & \qquad\nearrow &                & \searrow\qquad \\
				DC             & \longleftarrow & DD,             & CD             & \longleftarrow & DD
			\end{pmatrix} $\end{tabular}                           \\ \hline
		cycle-4 & $ \setlength{\tabcolsep}{1pt}
		\renewcommand{\arraystretch}{0.9} \begin{pmatrix}
			CC       & \rightarrow & CD         & CC & \rightarrow & DC \\
			\uparrow &             & \downarrow &
			\uparrow &             & \downarrow                         \\
			DC       & \leftarrow  & DD,        & CD & \leftarrow  & DD
		\end{pmatrix} $ \\ \hline
	\end{tabular}}
\end{table} 

\begin{figure}[htbp!]
\centering
\includegraphics[width=0.6\textwidth]{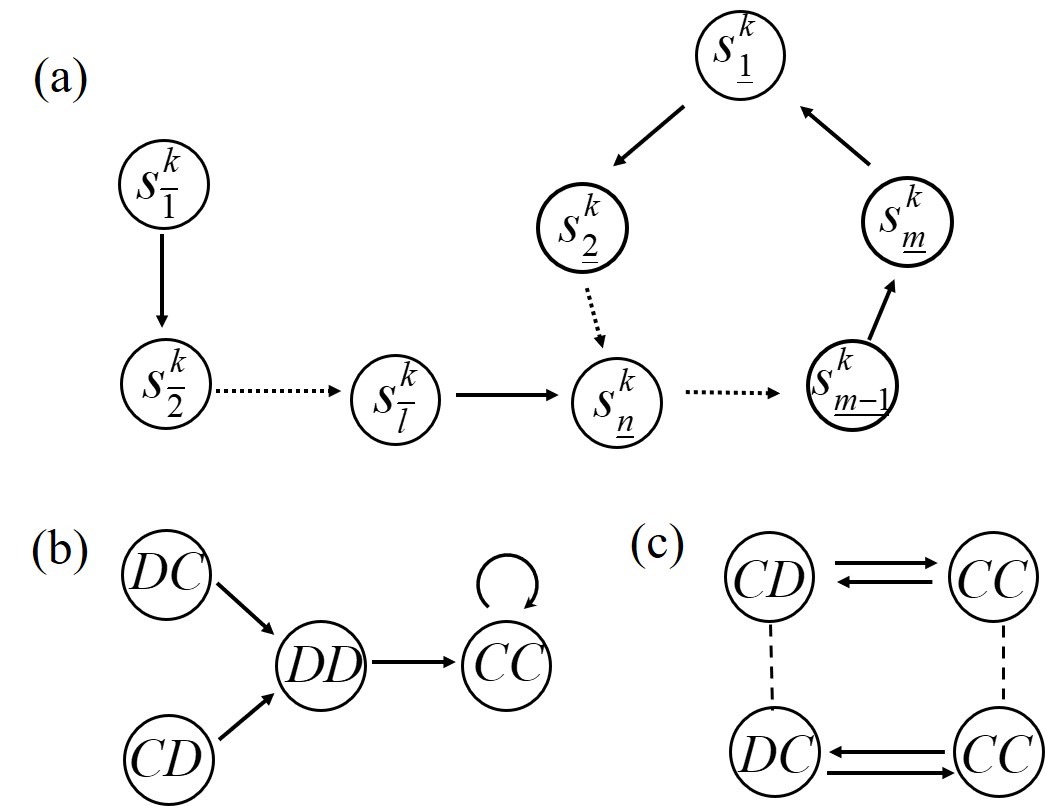}
\caption{ In (a), the diagram shows a cycle-$m$ mode and a \emph{key path}, along which the state transition will enter the mode. As an example of (a), (b) show the mode $\tuple{CC, CC}$ and the \emph{key paths} for $\tuple{\text{WSLS}, \text{WSLS}}$. In (c), we show the mode $\tuple{CC\leftrightarrow CD, CC\leftrightarrow DC}$ for Tolerance vs. Exploitation. 
} 
\label{fig:mode}  
\end{figure}

For a cycle-$1$ mode, the convergence rate of the Iris and Jerry's Q-tables is $1-\alpha+\alpha\gamma$, which increases with $\gamma$ but decreases with increasing $\alpha$. While,
for a cycle-$m$ mode with $m\geq 2$, the dynamics of Q-table for any agent $k\in\set{i,j}$
in a cycle is as follows 
\begin{eqnarray}
\left(\begin{array}{c}
	Q^{k}_{s_{\underline{1}},a_{\underline{1}}}(\tau+ m)\\
	Q^{k}_{s_{\underline{2}},a_{\underline{2}}}(\tau+ m)\\
	Q^{k}_{s_{\underline{3}},a_{\underline{3}}}(\tau+ m)\\
	\vdots\\
	Q^{k}_{s_{\underline{n}},a_{\underline{n}}}(\tau+ m)
\end{array}\right)&=&
\left(\begin{array}{ccccc}
	1 & 0 & 0 & \cdots & 0\\
	0 & 1 & 0 & \cdots & 0\\
	0 & 0 & 1 & \cdots & 0\\
	\vdots & \vdots & \vdots & \ddots & \vdots\\
	\alpha\gamma & 0 & 0 & \cdots & 1-\alpha
\end{array}\right)
\cdots
\left(\begin{array}{ccccc}
	1 & 0 & 0 & \cdots & 0\\
	0 & 1-\alpha & \alpha\gamma & \cdots & 0\\
	0 & 0 & 1 & \cdots & 0\\
	\vdots & \vdots & \vdots & \ddots & \vdots\\
	0 & 0 & 0 & \cdots & 1
\end{array}\right)\cdot \nonumber\\
& &\left(\begin{array}{ccccc}
	1-\alpha & \alpha\gamma & 0 & \cdots & 0\\
	0 & 1 & 0 & \cdots & 0\\
	0 & 0 & 1 & \cdots & 0\\
	\vdots & \vdots & \vdots & \ddots & \vdots\\
	0 & 0 & 0 & \cdots & 1
\end{array}\right)\cdot\left(\begin{array}{c}
	Q^{k}_{s_{\underline{1}},a_{\underline{1}}}(\tau)\\
	Q^{k}_{s_{\underline{2}},a_{\underline{2}}}(\tau)\\
	Q^{k}_{s_{\underline{3}},a_{\underline{3}}}(\tau)\\
	\vdots\\
	Q^{k}_{s_{\underline{m}},a_{\underline{m}}}(\tau)
\end{array}\right)
+\left(\begin{array}{c}
	\alpha\Pi_{s_{\underline 2}}\\
	\alpha\Pi_{s_{\underline 3}}\\
	\alpha\Pi_{s_{\underline 4}}\\
	\vdots\\
	\alpha\Pi_{s_{\underline 1}}+\alpha^{2}\gamma\Pi_{s_{\underline 2}}
\end{array}\right) \nonumber\\
& &=\left(\begin{array}{ccccc}
	1-\alpha & \alpha\gamma & 0 & \cdots & 0\\
	0 & 1-\alpha & \alpha\gamma & \cdots & 0\\
	0 & 0 & 1-\alpha  & \cdots & 0\\
	\vdots & \vdots & \vdots & \ddots & \vdots\\
	(1-\alpha)\alpha\gamma & \alpha^2\gamma^2 & 0 & \cdots & 1-\alpha
\end{array}\right)\cdot
\left(\begin{array}{c}
	Q^{k}_{s_{\underline 1},a_{\underline 1}}(\tau)\\
	Q^{k}_{s_{\underline 2},a_{\underline 2}}(\tau)\\
	Q^{k}_{s_{\underline 3},a_{\underline 3}}(\tau)\\
	\vdots\\
	Q^{k}_{s_{\underline m},a^{k}_{\underline m}}(\tau)
\end{array}\right)\nonumber\ \\
 & &+ \left(\begin{array}{c}
	\alpha\Pi_{s^{k}_{\underline 2}}\\
	\alpha\Pi_{s^{k}_{\underline 3}}\\
	\alpha\Pi_{s^{k}_{\underline 4}}\\
	\vdots\\
	\alpha\Pi^{k}_{s_{\underline 1}}+\alpha^{2}\gamma\Pi_{s_{\underline 2}}
\end{array}\right)
\label{eq:dynamics_Q}
\end{eqnarray}
and the corresponding state transitions are shown in Fig.~\ref{fig:mode}(a).
Here, $\mathcal{S}^{k}_{\mu} = \set{s_{\underline 1},\cdots, s_{\underline m}}$ is the set of $k$’s states in the mode, e.g., $k$'s sets and its co-player are $\set{CC, CD}$ and $\set{CC, DC}$ in $\tuple{CC\leftrightarrow CD, CC\leftrightarrow DC}$, respectively. 
In the mode, agents' Q-tables must meet the constraints
\begin{eqnarray}
\max\limits_{a'}\{Q^{k}_{s_{\underline n},a^{\prime}}\}=Q^{k}_{s_{\underline n},a_{\underline n}}, \forall k\in\{i,j\}, s_{\underline n}\in \mathcal{S}^{k}_{\mu}. 
\label{eq:constraint_mode}
\end{eqnarray}
Here, the constraints for a mode is increased with the length of the mode. So, the long modes are generally more fragile than the short as mentioned in Subsec.~\ref{subsec:analysis}.

The eigenvalues of the matrix of right-hand side in Eq.~(\ref{eq:dynamics_Q}) meets 
\begin{eqnarray}
(\lambda+\alpha-1)^m = (\alpha\gamma)^m\lambda,
\label{eq:eigenvalues}
\end{eqnarray}
which are the horizontal coordinates of the intersection of curves
$y(\lambda) = (\lambda+ \alpha - 1)^m$ and $y(\lambda) = (\alpha\gamma)^m\lambda$.
Then, the maximum real eigenvalue $\lambda_{\max}$ is greater than $1-\alpha$.

To investigate the effects of learning parameters on the convergence rate, 
we derive both sides of Eq.~(\ref{eq:eigenvalues}) for $\alpha$ and $\gamma$, and we
have
\begin{eqnarray}
& &m(\lambda_{\max}+ \alpha - 1)^{m-1}+m(\lambda_{\max}+ \alpha - 1)^{m-1}\frac{\partial \lambda_{\max}}{\partial \alpha}  = \nonumber\\
& &\left[m\frac{(\lambda_{\max}+ \alpha - 1)^m}{\alpha} +\frac{(\lambda_{\max}+ \alpha - 1)^m}{\lambda_{\max}}\frac{\partial \lambda_{\max}}{\partial \alpha}\right]\nonumber
\end{eqnarray}
and 
\begin{eqnarray}
& &m(\lambda_{\max}+ \alpha - 1)^{m-1}\frac{\partial \lambda_{\max}}{\partial \gamma} \nonumber\\
& &= \left[\frac{m(\lambda_{\max}+ \alpha - 1)^m}{\gamma} + \frac{(\lambda_{\max}+ \alpha - 1)^m}{\lambda_{\max}}\frac{\partial \lambda_{\max}}{\partial \gamma} \right]\nonumber.
\end{eqnarray}
According to the above equation, we get 
\begin{eqnarray}
\frac{\partial \lambda_{\max}}{\partial \alpha} = \frac{m\lambda_{\max}(1-\lambda_{\max})}{\alpha[(\alpha-1)+\lambda_{\max}(1-m)]}<0 
\label{eq:convergence_alpha}
\end{eqnarray} 
and 
\begin{eqnarray}
\frac{\partial \lambda_{\max}}{\partial \gamma} = \frac{\lambda_{\max}m(\lambda_{\max}+\alpha-1)}{\gamma[(1-\alpha)+\lambda_{\max}(m-1)]}>0 
\label{eq:convergence_gamma} 
\end{eqnarray} 
under $m\ge 2$ and $(1-\alpha)<\lambda_{\max}<1$. So, in any mode, 
the convergence rates of Q-tables for both agents increases with $\alpha$ but decreases of $\gamma$.

\newpage
\subsection{Stability of COMs/COPs}\label{subsec:ap_stableQ}

Any Q-value, $Q^{k}_{s_{\underline n},a_{\underline n}}$, in the mode of Fig.~\ref{fig:mode} will converge to fixed
\begin{eqnarray}
Q^{k*}_{s_{\underline n},a_{\underline n}} = \frac{\Pi_{s_{\underline{n+1}}}+\gamma \Pi_{s_{\underline {n+2}}}+\cdots \gamma^{m-l} \Pi_{s_{\underline m}} + \gamma^{m-l+1} \Pi_{s_{\underline 1}} + \cdots \gamma^{m-1} \Pi_{s_{\underline{n-1}}}}{1-\gamma^m}\label{eq:core_mode}
\end{eqnarray}
if the mode is a stable COM. It is obvious that the next state is determined by the current states if both agents' policies are remaining and $\epsilon\rightarrow 0$. In the case, the agents will enter the COM through determined state transition paths, called as \emph{key paths}. Then, the Q-values of corresponding policies of both agents are also converged along these paths. Without loss of generality, we only focus on the Q-values in one of key paths, $\mathcal{P}^{k_{\nu}}= \set{s_{\bar{1}}, s_{\bar{2}},\cdots, s_{\bar{l}}}$, as Fig.~\ref{fig:mode}(a) shown. The Q-values in $\mathcal{P}^{k_{\nu}}$ will also converge to fixed values if the Q-values are always satisfying following constraints
\begin{eqnarray}
\max\limits_{a'}\{Q^{k}_{s_{\overline n},a^{\prime}}\}=Q^{k}_{s_{\overline n},a_{\overline n}}, \forall k\in\{i,j\}, s_{\overline n}\in \mathcal{P}^{k}_{\nu},
\label{eq:constraint_path}
\end{eqnarray}
i.e., both agents policies on the path are unchanged.
According to Fig.~\ref{fig:mode}(a), we get the stable Q-values in the path as follows 
\begin{eqnarray}
Q^{k*}_{s_{\overline{i}},a_{\overline{i}}} = \Pi_{s_{\overline{i+1}}} + \gamma \Pi_{s_{\overline{i+2}}}+\cdots + \gamma^{l-i}\Pi_{s_{\overline{l}}}+\gamma^{l-i+1}Q^{*n}_{s_{\underline n},a_{\underline n}}.
\label{eq:core_policy}
\end{eqnarray}
Here, $s_{\underline n}$ is the first state in COM to pass along the path.
It is obvious that for a given exploration rate $\epsilon$ the convergence rate also increases with $\alpha$ but decreases with $\gamma$. Note that the Q-values in the modes evolve much faster than on the key paths to the mode, while the Q-values on the key paths evolve much faster than on the other paths. 
Each state on key paths or COM only has one state as the next transition. It means that the correlation between  consecutive states is positive if they are also consecutive states in a COM or a key path.
But $p(s)$ for $s$ in $\mathcal{S}^{k}_{\mu}$ is much greater than for $s$ in $\mathcal{P}^{k}_{\nu}$ because the only way to leave a stable COM is to explore (see Fig.~\ref{fig:state_transition}(a) and (c)).

In Fig.~\ref{fig:mode} (a), the COM will be broken as long as any constraint in Eq.~(\ref{eq:constraint_mode}) is unsatisfied. And the state transitions cannot return to the mode through $\mathcal{P}_{\mu}^{k}$ after leaving the mode by exploration, as long as the constraints in Eq.~(\ref{eq:constraint_path}) cannot be satisfied.
In both cases, at least one agent has changed its policy at some state and the COP is not unique.
Thereafter, the COP may switch by exploration.
This means that a COM might become unstable before the Q-tables of the COM converges to fixed.
Our analysis show that high $\alpha$ and low $\gamma$ could reduce robustness of both agents'  policies of COPs under competition and thus shorten characteristic time of the corresponding COMs. It is the reason why $f_c$ is more volatile for high $\alpha$ and low $\gamma$ as Fig.~\ref{fig:time_series} shows. Furthermore, cooperation also becomes fragile in the case when co-player's policy becomes unpredictable, because all-defection for any agent is the best policy. 
It is important to note that when the exploration rate is low, it is the exploration fluctuation, not the exploration itself, that weakens the robustness of the competing COPs. In contrast,
the exploration may enhance the robustness of the COPs because the exploration is benefit to maintain the constraints in Eqs.~(\ref{eq:constraint_mode}) and (\ref{eq:constraint_path})

\subsection{Supplementary Simulations for Tipping Points}\label{subsec:ap_tipping}

\begin{figure}[htbp!]
\centering
\includegraphics[width=0.75\textwidth]{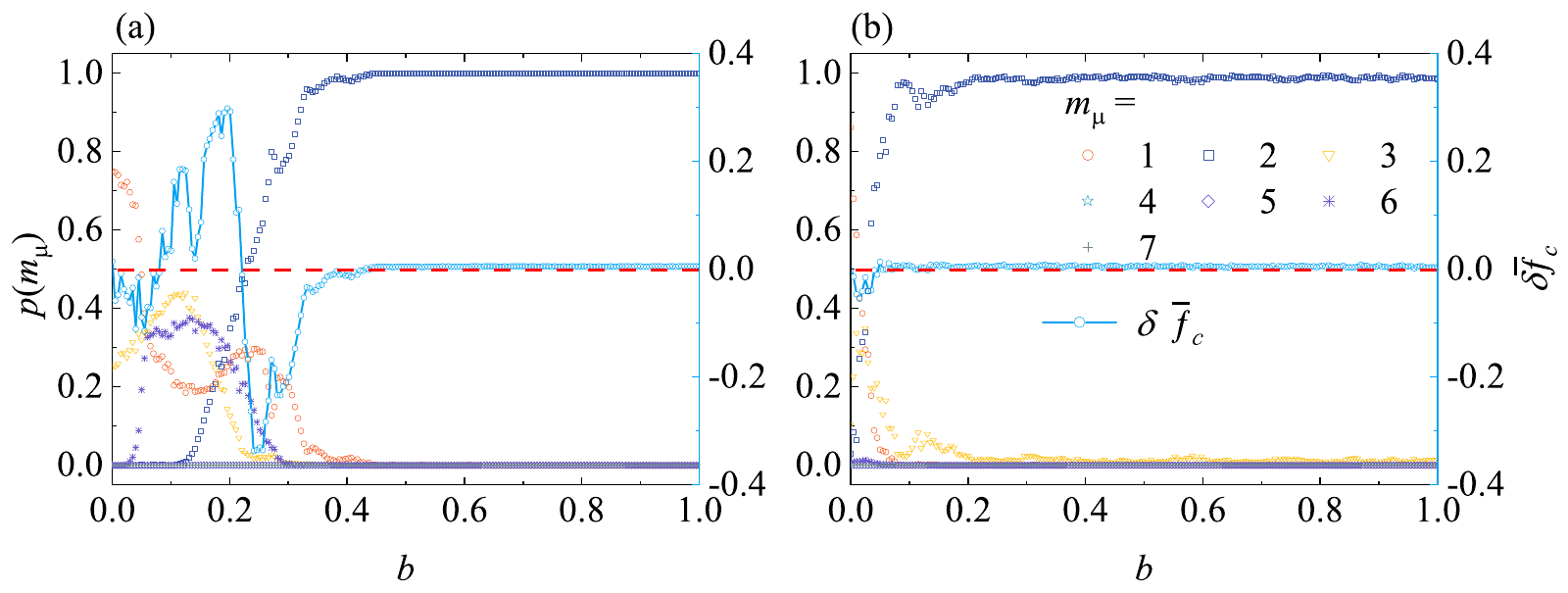}
\caption{{\bf Probabilities of different coupling-modes and the average cooperation preference gap with and without exploration.}
The coupling-modes are also corresponding to Fig.~\ref{fig:transition_modes}. In (a-b), the combinations of $(\alpha, \gamma) = (0.1, 0.1)$, $(0.1, 0.5)$. The red dashed line is $p(m_{\mu}) = 0.5$ or the gap $\delta \bar{f}_c = 0$ for comparison. As $\delta \bar{f}_c$ is above/below the line, it means that the exploration facilitates/suppresses the cooperation preference.
In the figure, exploration stops after $2\times 10^8$ rounds and the initial exploration rate is $\epsilon = 0.01$. The probabilities of different modes in Eq.~(\ref{eq:probabilities_mode}) is obtained from $2000$ runs.
}\label{fig:ap_pm}  
\end{figure}

According to Fig.~\ref{fig:probabilities_modes} in Sec.~\ref{subsec:exploration}, we conjecture that the tipping point for whether exploration can promote cooperation is always corresponding to $p(m_2) = 0.5$. To verify that, we give more 
results on $p(m_{\mu})$ and $\delta \bar{f}_c$ in Fig.~\ref{fig:ap_pm}. The results, especially (a), support our conjecture.   

\section{Cohen’s Kappa Coefficients for Temporal Correlations}\label{subsec:ap_kappa}

\begin{figure}[htbp!]
\centering
\includegraphics[width=0.75\textwidth]{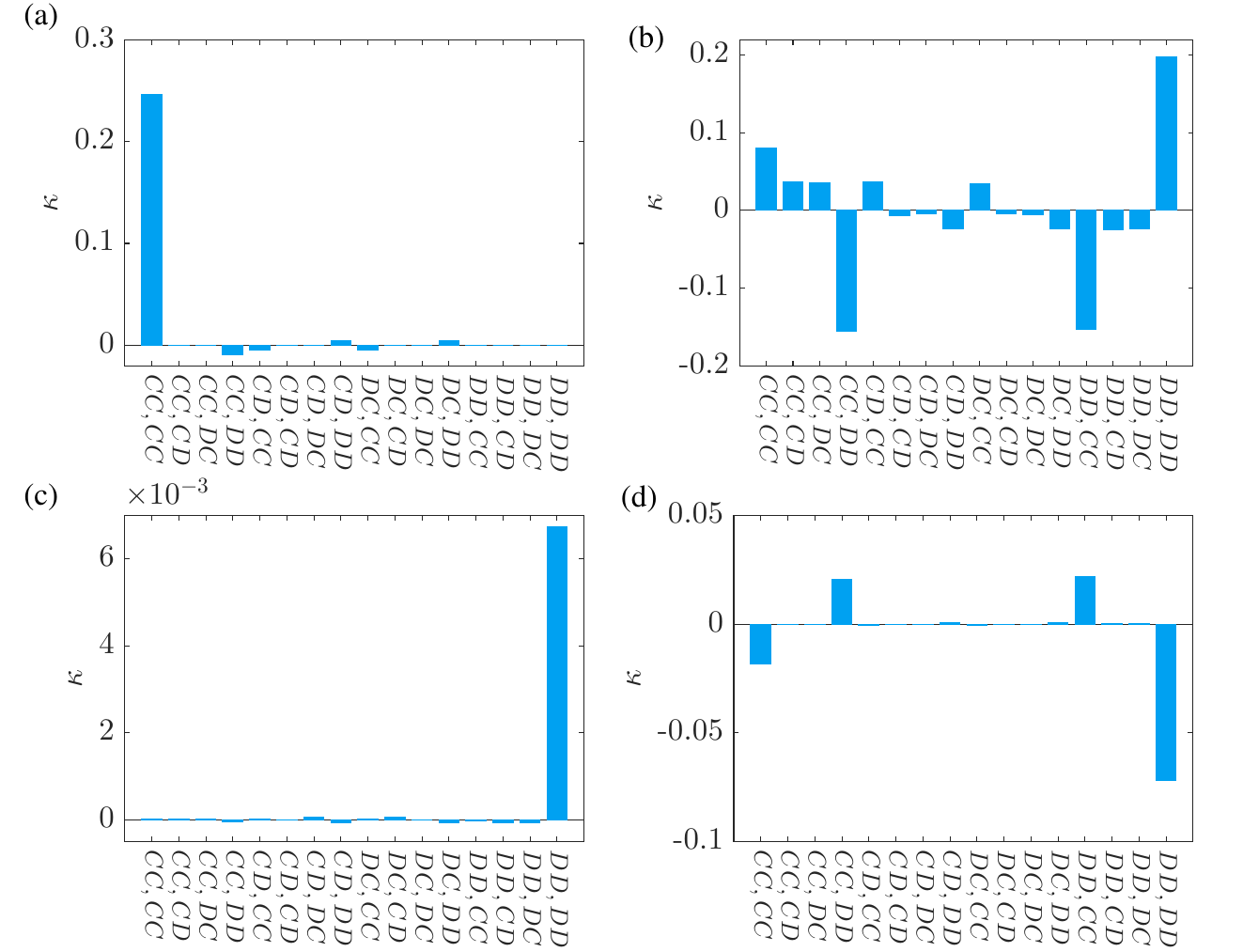}
\caption{{\bf The Cohen's kappa coefficients of consecutive states.} 
The learning parameter combinations in (a-d) are $(\alpha,\gamma) = (0.1, 0.9)$, $(0.5, 0.9)$, $(0.9, 0.5)$ and $(0.1, 0.1)$, respectively. (a-d) more clearly display the temporal correlations between consecutive states in different regions as shown. Positive kappa coefficients mean positive temporal correlation between the consecutive states.
Other parameters: $\epsilon = 0.01$, $b = 0.2$. The scale and location of the window are $T = 6\times 10^8$ and $t = 8\times 10^8$ after discarding $2\times 10^8$ rounds.
}
\label{fig:kappa}  
\end{figure}

In analysis of COM in Sec.~\ref{subsec:co_modes}, we give the gap between $p(s,s^{\prime})$ and $p(s)p(s^{\prime})$ to show the correlation between two consecutive states. But, the gap becomes invisible as $p(s,s^{\prime})$ is close to $0$ or $1$. So, we here employ the Cohen's kappa coefficients for better visualisation. The Cohen's kappa coefficient of two consecutive rounds is defined as 
\begin{center}
\begin{equation}
\kappa:=\frac{p(s,s^{\prime})-p(s)p(s^{\prime})}{1-p(s)p(s^{\prime})}.
\label{eq8}
\end{equation}
\par\end{center}
Fig.~\ref{fig:kappa} shows the results corresponding to Fig.~\ref{fig:p_p}. As Fig.~\ref{fig:kappa} (a)
and Fig.~\ref{fig:p_p} (a) show, the correlation between the consecutive $CC$s is positive as $CC$ is the dominant state for a low $\alpha$ and large $\gamma$. Similarly, the consecutive $DD$s also have a positive correlation when $DD$ is as the dominant state under high $\alpha$.  For the low $\alpha$ and 
$\gamma$, even though $DD$ is still the dominant state, the correlation between $DD$s is negative as Fig.~\ref{fig:kappa}(d) and Fig.~\ref{fig:p_p}(d) shown.  

\bibliographystyle{elsarticle-num-names} 
\bibliography{main}

\end{document}